\documentclass[fleqn,10pt]{wlscirep}
\usepackage{diagbox, multirow}
\usepackage[utf8]{inputenc}
\usepackage[T1]{fontenc}

\usepackage{siunitx}
\usepackage{amsmath}
\usepackage{float}
\restylefloat{table}
\usepackage{subcaption}
\usepackage{booktabs}
\usepackage{relsize}

\title{Impact of gas background on XFEL single-particle imaging} 

\author[1,+]{Tong You}
\author[2]{Johan Bielecki}
\author[1,3,*]{Filipe R.N.C. Maia}

\affil[1]{Laboratory of Molecular Biophysics, Institute for Cell and Molecular Biology, Uppsala University, Box 596, 75124 Uppsala, Sweden,}
\affil[2]{European XFEL, Holzkoppel 4, 22869 Schenefeld, Germany}
\affil[3]{NERSC, Lawrence Berkeley National Laboratory, Berkeley, California 94720, United States}
\affil[*]{filipe.maia@icm.uu.se}

\begin{abstract}
Single-particle imaging (SPI) using X-ray free-electron Lasers (XFELs) offers the potential to determine protein structures at high spatial and temporal resolutions without the need for crystallization or vitrification. However, the technique faces challenges due to weak diffraction signals from single proteins and significant background scattering from gases used for sample delivery. A recent observation of a diffraction pattern from an isolated GroEL protein complex \cite{ekeberg:2024} had similar numbers of signal and background photons. Ongoing efforts aim to reduce the background created by sample delivery, with one approach replacing most of the used gas with helium \cite{tej:2024}. 
In this study, we investigate the effects of a potentially reduced background on the resolution limits for SPI of isolated proteins under different experiment conditions. As a test case, we used GroEL, and we used experimentally measured parameters for our simulations.
We observe that background significantly impacts the achievable resolution, particularly when the signal strength is comparable to the background, and a background reduction would lead to a significant improvement in resolution.
\end{abstract}
\begin{document}

\flushbottom
\maketitle
\thispagestyle{empty}

\section*{Introduction}
Unlike traditional synchrotron X-ray sources, X-ray Free Electron Lasers (XFELs) provide orders-of-magnitude higher peak intensities and ultrashort pulse durations, typically on the femtosecond scale. Using them it may be to image individual biological particles to high spatial and temporal resolution, using a technique known as single-particle imaging (SPI) \cite{xfel:2010,bieleckispi:2020,sobolev:2020}. This enables imaging ultrafast dynamics and offers advantages over X-ray crystallography and cryo-electron microscopy (cryo-EM). In particular, SPI circumvents the need to crystallize the sample, as required in X-ray crystallography, and in principle allows much higher time resolution than cryo-EM.

The potential of SPI was first theorized by Neutze et al. \cite{neutze:2000}, who predicted that using sufficiently short XFEL pulses could allow proteins to diffract before the intense ionizing radiation destroys the sample. This "diffraction-before-destruction" concept was experimentally validated by Chapman et al. in 2006 \cite{chapman:2006}, demonstrating the feasibility of obtaining structural information from single particles before they are vaporized by the beam. 

However, significant challenges remain for SPI to achieve high-resolution reconstructions of single proteins \cite{aquila_2015,bielecki_2020}. One of the primary difficulties arises from the weak diffraction signal generated by single proteins, as SPI does not benefit from the signal enhancement provided by the numerous sample copies that make up the crystal lattice in X-ray crystallography. The protein also needs to be delivered into the XFEL beam. To address this, efficient sample delivery methods are critical. Various approaches, including gas dynamic virtual nozzles (GDVNs) and electrospray ionization (ESI), have been developed to introduce proteins into the XFEL beam. ESI, in particular, has been invaluable at delivering small samples as its small droplet sizes are essential to minimizes contaminants, which is common in GDVN-based delivery \cite{bielecki:2019}.

Despite these advances, SPI has so far achieved only two-dimensional reconstructions of cells \cite{hantke:2014,vdschot:2014} and three-dimensional reconstructions of viruses \cite{ekebergmimi:2015,lundolm:2018}. A complete three-dimensional reconstruction of a single protein remains elusive. Recently, the first diffraction pattern from a protein complex, GroEL, was recorded \cite{ekeberg:2024}. Although many patterns were collected, only one could confidently be attributed to a single GroEL protein complex, highlighting the difficulty of obtaining high-quality diffraction data from smaller particles.

Several factors contribute to the challenges of SPI with proteins. The smaller size of proteins inherently reduces the scattering signal compared to larger particles such as viruses. Additionally, background noise further complicates the reconstruction process. While there are numerous simulation studies of SPI \cite{poudyal_2020, nakano:2018, nakano:2023,Tegze:zf5017,donatelli_2017}, only a few incorporate external noise sources. While some have focused on the impact of detector noise \cite{juncheng:2022}, and other works have examined radiation damage \cite{yoon_2016,Fortmann-Grote:2017,stlin_2019,stransky:2024} and water layer effects \cite{stransky:2021,juncheng:2023}, the dominant issue remains the excess scattering from the background gas. Recent improvements in ESI, replacing most of the gas used with helium, have significantly reduced sample delivery related background scattering, potentially improving the signal-to-noise ratio for protein diffraction \cite{tej:2024}.

In this work, we investigate the effects of background scattering on the quality of reconstructed protein structures. Importantly, while many other simulations use X-ray fluence values assuming optimal performance of the X-ray optics and negligible radiation damage, we do not. Instead we use experimentally derived fluence values, based on the elastic scattering data from known samples. This way we implicitly include the effects of realistic optics and radiation damage effects. Furthermore we incorporated experimentally measured gas scattering, aiming to provide a realistic assessment of the achievable resolution in SPI experiments. Our results show that background scattering significantly influences the resolution of the reconstructions as well as the number of diffraction patterns required and hope these can help guide future experimental designs.

\section*{Results and Discussion}
We chose the \emph{Escherichia coli} chaperonin GroEL \cite{hayerhartl_2016} to look at the effects of experimental background on the resolution limits because it was the first isolated protein observed at an XFEL \cite{ekeberg:2024}, and background noise was a limiting factor in that experiment. GroEL has also been extensively characterized in past studies and many high-resolution X-ray structures are available.

We simulated diffraction patterns of randomly oriented single GroEL particles, from the PDB model 1SS8 \cite{charu:2000}, using experimentally measured parameters for the Small Quantum Systems (SQS) \cite{meyer:2020} and the Single Particles, Clusters, and Biomolecules \& Serial Femtosecond Crystallography (SPB/SFX) \cite{mancuso:2019} instruments of the European XFEL at 1.2, 2.5 and 6 keV. (see 2D pattern simulation in the Methods section for exact details).  

\begin{figure}[h]
\centering
\includegraphics[scale=1.0]{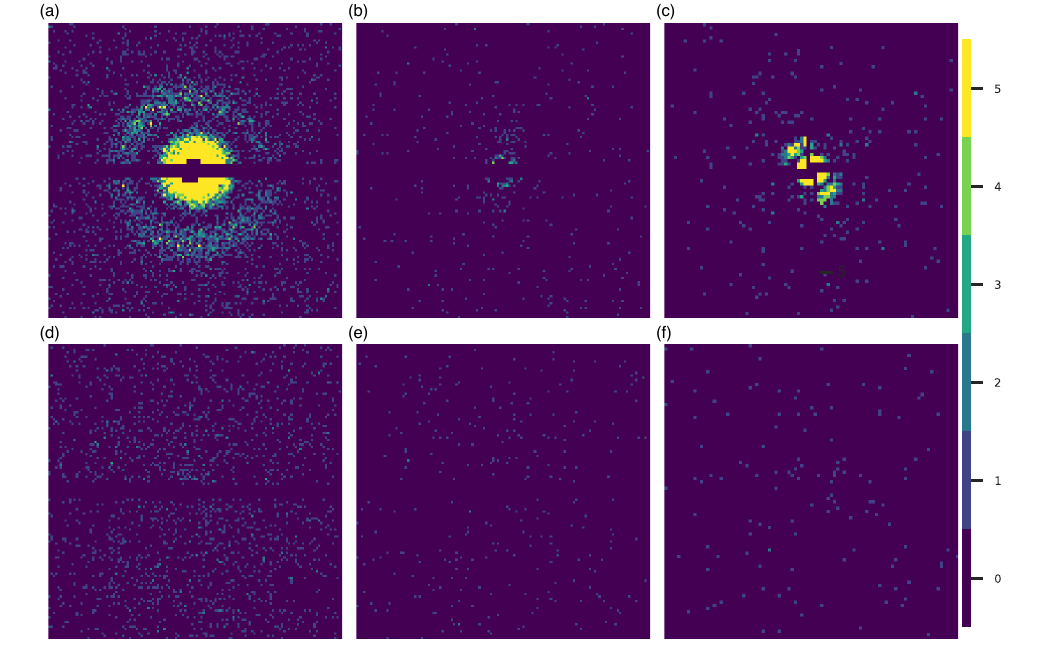}
\caption{Diffraction patterns under medium noise conditions. The top row shows protein diffraction combined with background at \textbf{(a)} 1.2 keV, \textbf{(b)} 2.5 keV, and \textbf{(c)} 6.0 keV. The bottom row shows background only scattering for \textbf{(d)} 1.2 keV, \textbf{(e)} 2.5 keV, and \textbf{(f)} 6.0 keV. The color bar denotes photon counts per pixel.}
\label{diffraction_pattern_panel}
\end{figure}

We combined the simulated patterns with experimentally measured background, yielding a diffraction pattern with experimental noise (see Fig. \ref{diffraction_pattern_panel} for examples). As experimental background was not available at 6 keV we modeled it based on the low energy background (see Background modeling in Methods). Given recent developments in sample delivery injection \cite{tej:2024}, the background in future experiments is expected to be significantly reduced. To investigate the consequences of this, we also simulated patterns where the background was reduced by a factor of 10 and 100 before incoherent addition as well as patterns without added background. The three background levels will be referred to as high, medium, low background and zero background respectively.

The number of recorded diffraction patterns is also an important parameter since averaging techniques can boost the signal-to-noise ratio. We tested datasets with $10^3$, $10^4$, and $10^5$ patterns. As the results for $10^4$ fall in between the other two datasets, we shown them only in the Supplementary Information. 

{
\begin{table}[h]
\centering
\begin{tabular}{lccc}
\toprule
& 1.2 keV  & 2.5 keV  & 6.0 keV  \\
\midrule
GroEL signal & 5644 & 101 & 1048 \\
High background & 13967 & 2235 & 1044 \\
Medium background & 1397 & 223 & 104 \\
Low background & 140 & 22 & - \\

\bottomrule
\end{tabular}
\caption{Average photon counts for the signal expected from a single GroEL particle along with the background for each level investigated.
}
\label{noise_level_photon_count}
\end{table}
}
Since signal-to-noise is an important aspect of the present study, Table \ref{noise_level_photon_count} shows a comparison of the average number of photons for a single GroEL particle with the background at the different levels investigated.

To orient the 2D patterns into a consistent 3D volume we used the Dragonfly \cite{kartik:2016} package which employs the Expand-Maximize-Compress (EMC) algorithm \cite{emc:veit}.
In principle one can perform background correction either before or after the EMC assembly. We found that correcting the 2D diffraction patterns leads to instabilities in the EMC assembly and failure to converge. Instead, we did two separate EMC assemblies, one of the background-corrupted protein diffraction signal and another one of only the background diffraction which we then subtracted from the first one.

\begin{figure}[H]
\centering
\includegraphics[scale=1.0]{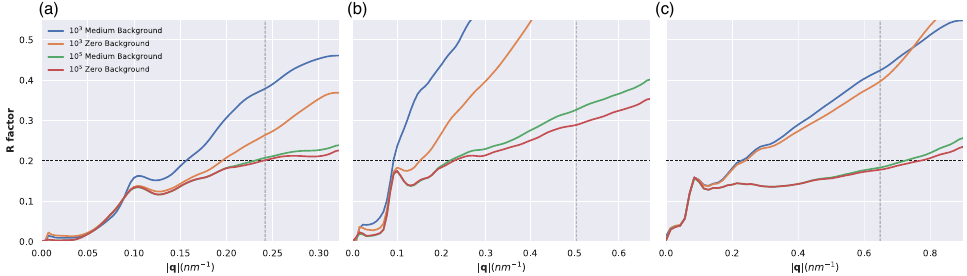}
\caption{R factor curves for all three geometries under medium and zero background conditions for \textbf{(a)} 1.2 keV, \textbf{(b)} 2.5 keV and \textbf{(c)} 6.0 keV. The horizontal dashed line corresponds to the 0.2 threshold used to define the resolution.}
\label{r_factor_all_medium} 
\end{figure}

To monitor the quality of the assembled 3D intensities we used the R-factor \cite{rfactor:2005}, between the assembled intensities the scattering factors calculated from the PDB model. We checked the resolution at which it drops below 0.2 and denominate this R-factor resolution. For all three energies, the R-factor resolution improves as the number of patterns increases, or when noise decreases. 
In Figure \ref{r_factor_all_medium}, the zero and medium background R-factor curves are shown for different dataset sizes.
The background impact on the EMC intensity assembly is largest when only $10^3$ patterns are used. With few patterns, the R-factor increases substantially at high-q and the resulting assembled intensities have less well-resolved speckles (see Fig X (central slices) in the SI).

The recovered 3D intensity volumes were then phased with libspimage \cite{hawk:2010} (see methods for more details). We estimated the quality of the reconstructed structures using the Phase Retrieval Transfer Function (PRTF) \cite{prtf:2006} and the Fourier Shell Correlation \cite{fsc:1} (FSC). 

\begin{figure}[h]
\centering
\includegraphics[scale=1.0]{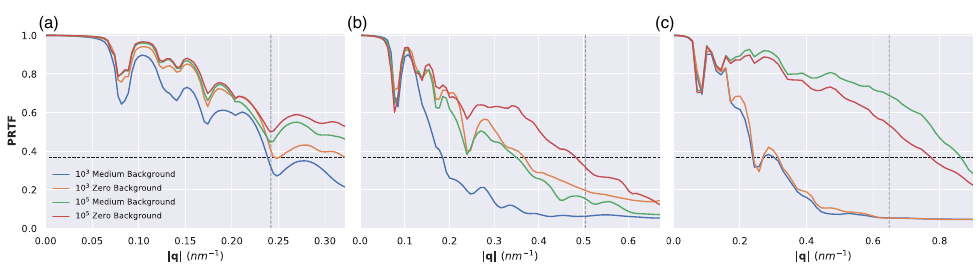}
\caption{PRTF curves for all three geometries under medium and zero background conditions for \textbf{(a)} 1.2 keV, \textbf{(b)} 2.5 keV and \textbf{(c)} 6.0 keV. The horizontal dashed line corresponds to the 1/e threshold traditionally used to define the resolution. The vertical dashed line indicates the edge resolution for each geometry.}
\label{prtf_all_medium}
\end{figure}

The PRTF curves for different conditions are shown in Figure \ref{prtf_all_medium}.
There is a large gap between the PRTF curves for $10^3$ and $10^5$ datasets at 6 keV. The difference due to the number of patterns is less pronounced at 1.2 and 2.5 keV. 
There is also very little difference between the dataset with background and without for the 6 keV case. In this case the resolution is limited by the number of patterns collected.

In Figure \ref{prtf_all_medium}, for $10^5$ 6 keV the resolution of the medium background reconstruction is better than the zero background one, although the difference is very small. This counterintuitive result is probably due to the residual noise helping the phasing algorithm explore a larger phase space. This discrepancy is not visible in the R-factor curves (see Figure \ref{r_factor_all_medium}), which are unaffected by phasing. 

\begin{table}[h]
\centering
\begin{tabular}{rrrrrr}
\multicolumn{6}{c}{Resolutions in nm} \\
\toprule
Photon Energy & Dataset Size & High & Medium & Low  & Zero Background\\
\midrule
1.2 keV & $10^3$ & 6.9 / 4.7 / 3.5 & 6.4 / 3.9 / 3.2 & 5.2 / 3.1 / 3.1 & 5.1 / 3.1 / 3.2 \\
1.2 keV & $10^5$ & 5.2 / 4.1 / 3.2 & 4.3 / 3.1 / 3.1 & 4.6 / 3.1 / 3.1 & 4.2 / 3.1 / 3.1 \\ 
\midrule
2.5 keV & $10^3$ & - & 10.9 / 5.3 / 6.2 & 11.1 / 5.0 / 4.8 & 6.6 / 2.8 / 2.7 \\
2.5 keV & $10^5$ & 6.7 / 4.5 / 2.9 & 4.6 / 2.8 / 2.3 & 4.7 / 2.0 / 1.6 & 4.4 / 2.0 / 1.5 \\
\midrule
6.0 keV & $10^3$ & 5.1 / 4.2 / 2.9 & 4.4 / 3.3 / 2.7 & - & 4.2 / 3.2 / 2.1 \\
6.0 keV & $10^5$ & 1.7 / 1.4 / 1.1 & 1.4 / 1.2 / 1.1 & - & 1.3 / 1.3 / 1.1 \\
\bottomrule
\end{tabular}
\caption{Comparison of multiple resolution metrics for all $10^3$ and $10^5$ datasets. For each condition, the R-factor resolution is given, followed by the PRTF and finally the FSC, the three values separated by forward slashes. For a precise description of how resolutions are calculated, refer to the Methods section.}
\label{all_resolutions}
\end{table}

Table \ref{all_resolutions} shows multiple resolution metrics for the tested conditions. Most reconstructions converged, except those at 2.5 keV with less than $10^5$ patterns under the high noise condition. Interestingly in almost all cases, the FSC is better than the resolution as determined by the PRTF. Also of note, the 0.2 threshold which we used for our R-factor resolution leads to a pessimistic estimate of the resolution compared to the other metrics. 

\begin{figure}[h]
\centering
\includegraphics[scale=1.0]{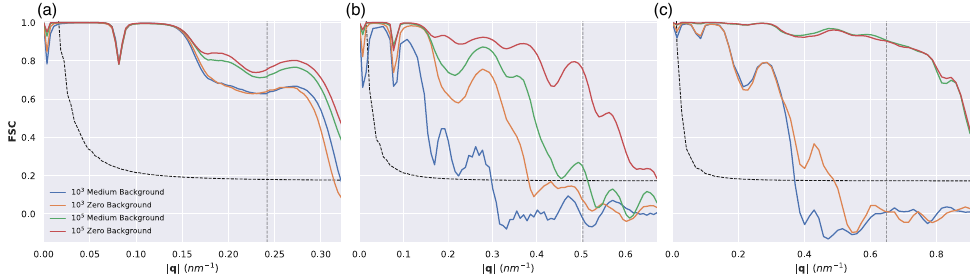}
\caption{Radially averaged FSC curves under medium and zero background conditions of electron density reconstructions for \textbf{(a)} 1.2 keV, \textbf{(b)} 2.5 keV and \textbf{(c)} 6 keV. The half-bit curve is shown as a black dashed line.}
\label{fsc_all_medium}
\end{figure}

The FSC is the most common method to assess a reconstruction in cryo-EM. Since there is a ground-truth structure available, there is no requirement to separate the experimental data into two separate half-data sets and instead we can compare directly to the model density.
FSC curves are shown in Figure \ref{fsc_all_medium} (see the SI for other conditions).

The number of patterns seems to make the most difference for the 6 keV dataset, while at 1.2 keV there is little difference in the FSC as a function of the number of patterns as we are limited by the maximum scattering vector captured by the detector.
The zero and medium background reconstructions at 6 keV are indistinguishable for $10^5$ patterns as the FSC curves nearly overlap over the entire q-range. 

\begin{figure}[H]
\centering
\includegraphics[scale=0.43]{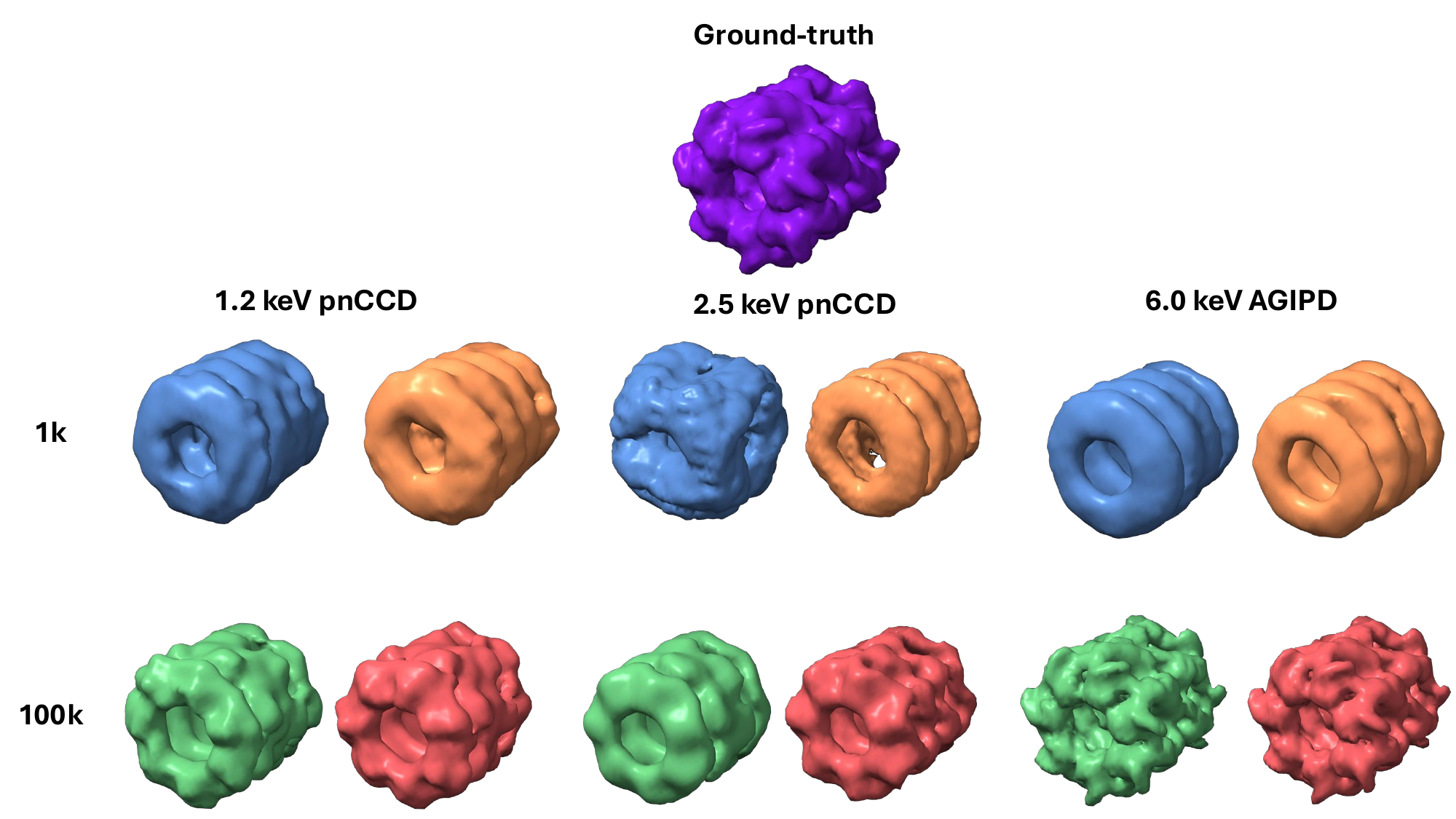}
\caption{Reconstructed electron density for all three energies for $10^3$, and $10^5$ pattern reconstructions. For each energy, the left column (in blue and green) represents the medium background condition, and the right column (in orange and red) represents the zero background condition. The ground truth is shown on the top row.}
\label{dens_medium}
\end{figure}

It is also important to look at the reconstructions visually to assess their quality more qualitatively (see Figure \ref{dens_medium}).
The FSC curves in Figure \ref{fsc_all_medium} are all quite similar, but there is a significant difference in the details visible between the 1.2 keV $10^3$ and $10^5$ pattern reconstructions.
There is a large similarity between the background-corrupted and zero background reconstructions, which is consistent with the similarities observed in the FSC curves. 
There are however some departures from this observation. For example in Figure \ref{dens_medium} for 2.5 keV and $10^3$ patterns, the background-corrupted density is clearly worse than the Poisson-only reconstruction. The same thing is true for $10^5$ patterns, albeit to a lesser extent. This is also reflected in a gap between the corresponding FSC curves (Figure \ref{fsc_all_medium}). 

Overall we find that, for the conditions tested, the best resolution of around 1 nm is achieved using 6 keV with a nanofocus beam and $10^5$ patterns. A reduction in background by a factor of 10 (medium background) makes it possible to achieve roughly the same resolution with 10 times less patterns. However, it should be noted that achieving large dataset sizes with the nanofocus is difficult due to the small X-ray cross-section which lowers the number of particles which are imaged, which is why most studies to date have been done at lower energies. Even for soft X-rays, a reduction in background significantly improves the resolution and, for example in the case of the smaller datasets at 2.5 keV, can be the difference between a failed and a successful reconstruction.

\section*{Methods}
\subsection*{2D pattern simulation}
Diffraction patterns were simulated with Condor \cite{hantke:2016} with the parameters shown in Table \ref{simulation_parameters}. The orientation of each particle was randomized and the horizontal polarization of the EuXFEL beam was taken into account.

{
\begin{table}[h]
\centering
\begin{tabular}{lccc}
\toprule
& 1.2 keV & 2.5 keV & 6.0 keV \\
\midrule
Edge resolution (nm) & 4.13 & 1.98 & 1.51 \\
Corner resolution (nm) & 2.99 & 1.43 & 1.07 \\
Pixel size (\textmu m) & 75 & 75 & 200\\
Number of pixels & $ 1024 \times 1024$ & $1024 \times 1024$ & $1256 \times 1092$ \\
Downsampled pixels & $ 128 \times 128$ & $128 \times 128$ & $91 \times 91$ \\
Oversampling & 17.2 & 8.27 & 4.59 \\
Fluence (photons/\textmu m\textsuperscript{2}) & $ 1.46\times10^{12} $ & $ 2.33\times10^{11} $ & $ 1.04\times10^{13} $ \\ 
Fluence (\textmu J/\textmu m\textsuperscript{2}) & $ 280 $ & $ 93 $ & $ 10000 $ \\ 

\bottomrule
\end{tabular}
\caption{Key simulation parameters for each geometry investigated. The photon fluence scale covers three orders-of-magnitude. The detector parameters for the 1.2 and 2.5 keV correspond to a pnCCD \cite{pnccd:2021}, one of the detectors available at SQS, binned $8 \times 8$. For 6 keV they correspond to an AGIPD \cite{agipd:2019}, the main detector at SQB/SFX, cropped to a square and binned $12 \times 12$. Downsampling was done to reduce computational time while maintaining sufficient oversampling.}
\label{simulation_parameters}
\end{table}
}

A 1024x1024 pixels pnCCD detector \cite{pnccd:2021} with 75 \textmu m pixel size and 0.15 m distance to the interaction point was simulated with photon energies of 1.2 and 2.5 keV. Since successful phase retrieval only requires oversampling larger than 2, the computational load was reduced by downsampling the detector to 128 by 128 pixels. The 1.2 (2.5) keV simulations have an edge resolution of 4.13 (1.98) nm and a maximum corner resolution of 2.99 (1.43) nm. The 1.2 keV fluence was taken from previous experiments \cite{ekeberg:2024} to be approximately 280 \textmu J/\textmu m\textsuperscript{2}, while the 2.5 keV fluence was estimated at 93 \textmu J/\textmu m\textsuperscript{2}, taking into account the performance of the SQS beamline.

A 1256x1092 Adaptive Gain Integrating Pixel Detector (AGIPD)\cite{agipd:2019}, with 200 \textmu m pixels was simulated using a photon energy of 6.0 keV and fluence of 10 m\textmu J/\textmu m\textsuperscript{2}, which matches the fluence observed with the nano-KB mirrors \cite{bean_2016}. The simulated patterns were cropped to a square and 12x12 resulting in 91x91 pixels with an edge resolution of 1.51 nm and a corner resolution of 1.07 nm.

\subsection*{Background modeling \label{sec:Background modeling}}
In an SPI experiment using an aerodynamic lens-stack, the background from elastic scattering on gas consists of two components. The first component, $I_{\mathrm{line}}$ results from the scattering of gas at relatively uniform pressure $p$ extending from the interaction point to the detector. The second contribution, $I_{\mathrm{jet}}$ comes from the scattering of the higher-pressure gas jet immediately under the outlet of the aerodynamic lens. 

We describe the rotationally averaged scattering arising from a gas molecule by the Debye scattering formula, using $q=2 \sin{\theta}/\lambda$, where $\theta$ is half the scattering angle.
\begin{equation}
I(q) = \mathlarger{\mathlarger{\sum}}_i \mathlarger{\mathlarger{\sum}}_j f_i f^*_j \frac{\sin(2 \pi q d_{ij})}{2 \pi q d_{ij}},
\end{equation}
where $i,j$ goes over all atoms in the molecule, $f_i$ is the atomic form factor for atom $i$, and  $d_{ij}$ is the distance between atom $i$ and $j$. The range of scattering vectors covered by the detector varies as a function of the longitudinal distance $l$ between the gas molecule and detector plane. For a detector pixel with physical coordinates $(x,y)$, we have $\theta = \frac{\arctan{(r/l)}}{2}$ where $r=\sqrt{x^2+y^2}$. Using this explicit expression for $\theta$ we define $I(r,l,\lambda)$ as the scattering from a gas molecule at a distance $l$ from the detector towards a pixel located at radius $r$.

Hence, 
\begin{equation}
I_{\mathrm{line}}(r,\lambda)=N_{ph}n r_0^2\Omega\int_0^l I(r,l,\lambda)dl
\end{equation}
where $\Omega$ is the solid angle of the pixel,  $r_0$ is the classical electron radius, $n$ is the number of gas molecules per m$^3$, known as the volumetric number density, and $N_{ph}$ is the number of photons in the XFEL pulse.

Analogously we model $I_{jet}$ as:

\begin{equation}
I_\mathrm{jet}(r,\lambda)=N_{ph}n_{jet} r_0^2\Omega\int_{l-\Delta l/2}^{l+\Delta l/2} I(r,l,\lambda)dl
\end{equation}
where $\Delta l$ is the gas jet width and $n_{jet}$ its volumetric number density. 

The volumetric number density of background gas can be well approximated by the ideal gas law, $n=p/kT$,  while $n_{jet}$ is related to the flux of gas molecules through the aerodynamic lens. The total gas background can be expressed as $I_{gas}=I_{line} + I_{jet}$, with $n_{jet} = c \cdot n$, where $c$  equals the relative pressure increase in the gas jet relative the overall chamber pressure. $\Delta l$ is typically not known. However, the $q$ dependence of  $I_{\mathrm{jet}}$ is independent of $\Delta l$ as long as  $\Delta l \ll l$. It is thus possible to assume a small $\Delta l$ value and fit $c$ to a measured gas background. As $c$ only depends on the gas flow out of the aerodynamic lens, one such fit is sufficient to determine the gas background for different $\lambda$ and $l$ values. Here we use a gas background measured at 1.2 keV, and obtain $c=1950$. 

\subsection*{3D intensity assembly}
Dragonfly \cite{kartik:2016} was used to assemble the diffraction patterns. Rotational sampling was optimized to ensure that the average number of patterns per orientation was close to one. For $10^3$, $10^4$, and $10^5$ patterns, n=3 (1380), n=6 (10680), and n=12 (86520) rotational samples were used, respectively. In cases where high background noise relative to the protein signal led to convergence issues (i.e., pattern collapse into a few orientations), the rotational sampling was reduced to prevent this collapse and ensure proper reconstruction. Fluence scaling was enabled for all EMC reconstructions, as during an experiment the fluence on the sample is unknown. The recovered fluence was almost identical for all patterns, as expected. Due to the instability that may occur in high signal, high background, high rotational sampling, or low pattern reconstructions the deterministic annealing variant of EMC was utilized, with the annealing parameter $\beta$. This $\beta$ was gradually raised to 1.0 over the assembly. For high signal datasets (the $10^5$ with background and zero background) the initial $\beta$ was 0.001. For all the other reconstructions it was 0.01. The assembly was monitored until the mutual information and average log-likelihood plateaued at which point the iterations were stopped. The total number of iterations ranged from 180 to 650. 

\subsection*{3D phase retrieval}
Three-dimensional phase reconstructions were performed using libspimage \cite{hawk:2010}. Each reconstruction consisted of 500 Relaxed Averaging Alternating Reflections (RAAR) \cite{raar:2004} iterations followed by 450 Error Reduction (ER) \cite{fienup:1982}iterations. The RAAR feedback parameter $\beta$ was adjusted based on noise level: for 6.0 keV data, $\beta$ was set from 0.80 at the start to 0.85 at the end of reconstruction, for 2.5 keV at high and median background it went from 0.70 to 0.75 and for all other reconstruction it went from 0.6 to 0.65. The support was updated every 20 iterations using a volume-constraining version of the shrinkwrap algorithm \cite{shrinkwrap:2003} (where the total support volume is enforced at every support update iteration) with Gaussian blurring with standard deviations decreasing from 1.5 to 1.0 voxels. The initial estimate for the volume of the support was 2380 nm$^3$ and this was reduced to 2190 nm$^3$ at the end of the reconstruction.

500 independent reconstructions were performed for each dataset, and the 450 with the lowest real-space error were selected for alignment and averaging. The Phase Retrieval Transfer Function (PRTF) \cite{prtf:2006} was used to assess reproducibility and determine spatial resolution by calculating the intersection of the radial averaged of the critically
sampled PRTF \cite{lundolm:2018} (the oversampled PRTF convoluted with the Fourier transform the convex hull of the support)  with the $1/e$ threshold. Reconstructions were performed with the negative voxels in Fourier space free to take any intensity value.

\subsection*{3D structure alignment}
To calculate the FSC between our reconstructions and the ground truth, the 1SS8 PDB model, we have to rotationally and translationally align them. This was done using an automated script in ChimeraX \cite{chimerax:2023}. An electron density at a resolution of 15 Å was generated from the PDB using the molmap command in ChimeraX. Both the PDB-generated electron density and reconstructed electron densities were centered. Then 30 iterations of translational alignment were performed, followed by 30 iterations of rotational alignment. After the structures were aligned, the FSC was calculated by radially averaging the 3D correlation map using 1 voxel thick spherical shells. The resolution was determined by calculating the first intersection of the FSC with the (modified) half-bit threshold curve \cite{halfbit:2005}. The above procedure was repeated for the centrosymmetric version of each reconstruction, due to the handedness ambiguity inherent in all phasing procedures, and the one with the best FSC was kept. 

\subsection*{3D EMC model alignment}
3D scattering intensities were generated from the PDB model of GroEL using Condor.
The EMC assembled intensities were aligned to this ideal model intensities using ChimeraX and R-factors were calculated. We calculate the R factor until and including a given resolution \cite{rfactor:2005}. To determine the R-factor resolution, we find the first intersection with a constant threshold of 0.2. 

\section*{Conclusion}
In this study, we demonstrated that background noise significantly influences the achievable resolution in our imaging system. Using a 6 keV nano-focus beam and collecting $10^5$ patterns, we achieved the best resolution of approximately 1 nm. However, collecting large datasets using the nano-focus beam presents challenges due to the small X-ray cross-section, which limits the number of particles that can be effectively imaged. Notably, when the background noise was reduced by an order of magnitude, comparable resolution could be obtained with only one-tenth of the patterns, highlighting the crucial role of noise reduction in experimental efficiency.  While background reduction generally yielded substantial improvements in resolution, this effect was less pronounced at lower energies. Looking forward, achieving sub-nanometer resolution for biological samples will require further technical advances, particularly in reducing the size of the X-ray focal spots, increasing fluence, and improving sample delivery precision to fully utilize the smaller beam sizes.

\bibliography{full_reference_list}

\section*{Acknowledgements}
This work is supported by the Swedish Research Council (2018-00234 and 2019-06092) and the European Research Council (ERC Consolidator Grant 101088426).

\section*{Author contributions}
The concept of the paper was proposed by F.R.N.C.M.; T.Y. performed the reconstructions and analyzed the results; J.B. modeled the gas background scattering; T.Y. wrote the first draft of the manuscript. All authors discussed and interpreted the data, and contributed to writing the manuscript.

\newpage
\begin{center}
\LARGE{\textbf{Supplementary Information}}
\Large{
\\ for \\
\textbf{Impact of gas background on XFEL single-particle imaging}
\\ by \\
\textbf{T. You, J. Bielecki and F. R. N. C. Maia}}
\end{center}

\renewcommand\thefigure{S\arabic{figure}}
\renewcommand\thetable{S\arabic{table}}    

\begin{table}[h]
\centering
\begin{tabular}{rrrrrr}
\multicolumn{6}{c}{Resolutions in nm} \\
\toprule
Photon Energy & Dataset Size & High & Medium & Low  & Zero Background\\
\midrule
1.2 keV & $10^3$ & 6.9 / 4.7 / 3.5 & 6.4 / 3.9 / 3.2 & 5.2 / 3.1 / 3.1 & 5.1 / 3.1 / 3.2 \\
1.2 keV & $10^4$ & 5.5 / 4.2 / 3.3 & 5.0 / 3.3 / 3.1 & 4.4 / 3.1 / 3.1 & 5.1 / 3.1 / 3.1 \\
1.2 keV & $10^5$ & 5.2 / 4.1 / 3.2 & 4.3 / 3.1 / 3.1 & 4.6 / 3.1 / 3.1 & 4.2 / 3.1 / 3.1 \\ 
\midrule
2.5 keV & $10^3$ & - & 10.9 / 5.3 / 6.2 & 11.1 / 5.0 / 4.8 & 6.6 / 2.8 / 2.7 \\
2.5 keV & $10^4$ & - & 5.8 / 3.8 / 2.8 & 5.3 / 3.9 / 3.0 & 5.1 / 3.1 / 2.7 \\
2.5 keV & $10^5$ & 6.7 / 4.5 / 2.9 & 4.6 / 2.8 / 2.3 & 4.7 / 2.0 / 1.6 & 4.4 / 2.0 / 1.5 \\
\midrule
6.0 keV & $10^3$ & 5.1 / 4.2 / 2.9 & 4.4 / 3.3 / 2.7 & - & 4.2 / 3.2 / 2.1 \\
6.0 keV & $10^4$ & 4.6 / 2.4 / 1.9 & 4.8 / 1.7 / 1.2 & - & 4.8 / 1.5 / 1.1 \\
6.0 keV & $10^5$ & 1.7 / 1.4 / 1.1 & 1.4 / 1.2 / 1.1 & - & 1.3 / 1.3 / 1.1 \\
\bottomrule
\end{tabular}
\caption{Comparison of multiple resolution metrics for all tested conditions. For each condition, the R-factor resolution is given, followed by the PRTF and finally the FSC, the three values separated by forward slashes. For a precise description of how resolutions are calculated, refer to the Methods section.}
\label{all_resolutions}
\end{table}

\begin{figure}[h]
    \centering
    \begin{subfigure}[b]{0.35\textwidth}
    \caption{\textbf{1.2 keV pnCCD}}
    \includegraphics[scale=0.40]{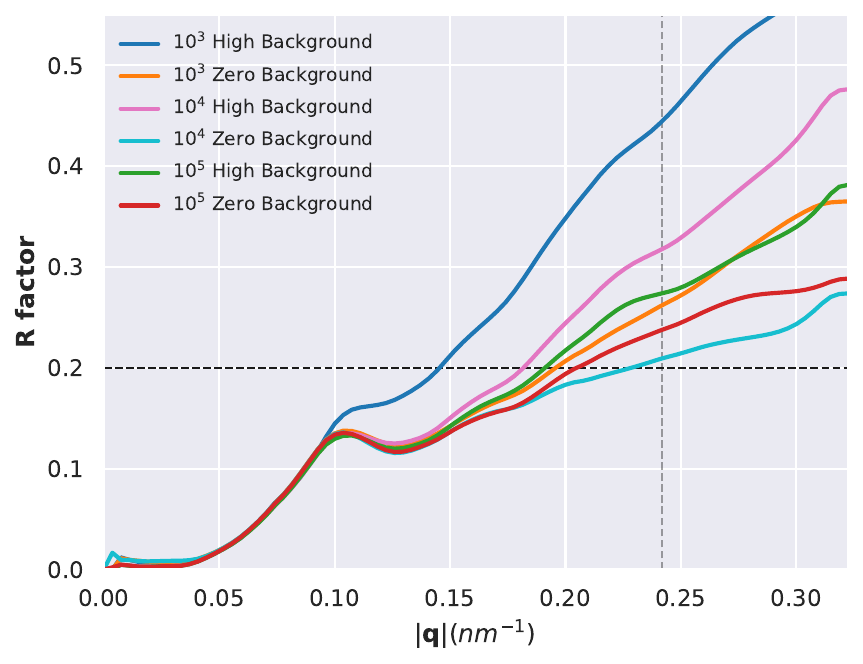}
    \label{rfactor_pnccd_1_normal}
    \end{subfigure}
    \centering
    \begin{subfigure}[b]{0.35\textwidth}
    \caption{\textbf{2.5 keV pnCCD}}
    \includegraphics[scale=0.40]{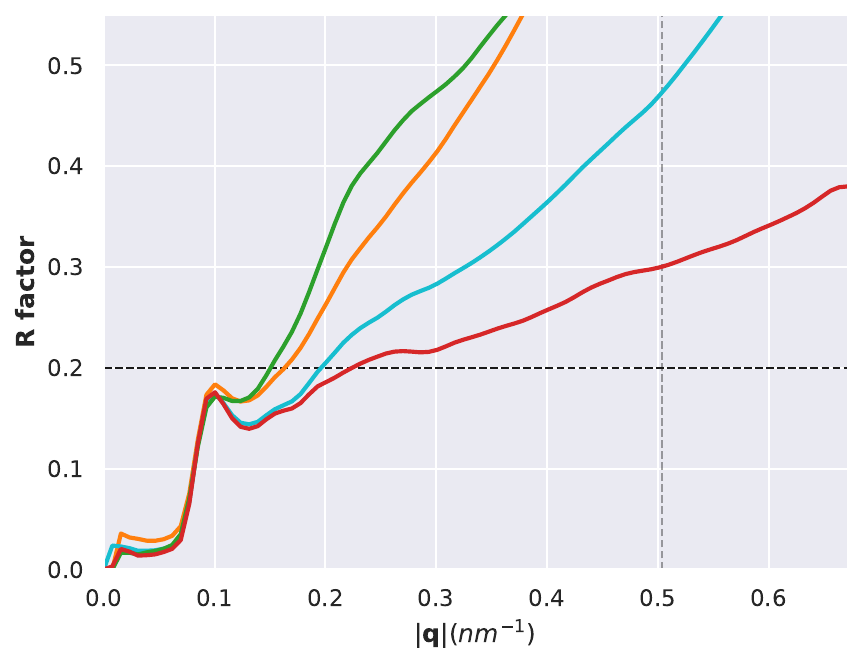}
    \label{rfactor_pnccd_2_normal}
    \end{subfigure}
    \centering
    \begin{subfigure}[b]{0.35\textwidth}
    \caption{\textbf{6.0 keV AGIPD}}
    \includegraphics[scale=0.40]{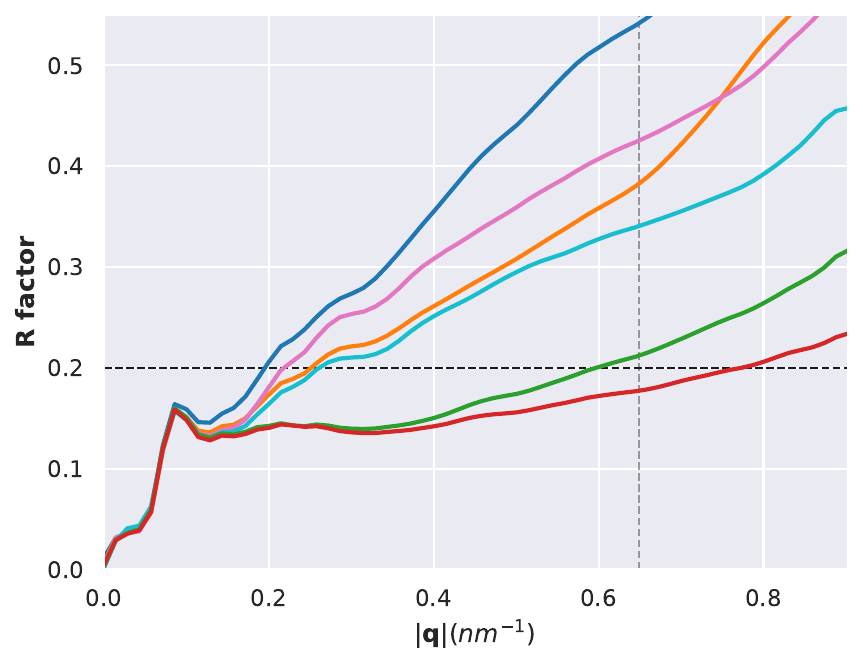}
    \label{rfactor_agipd_normal}
    \end{subfigure}
    \caption{R factor curves for all three energies under high and zero background conditions.}
    \label{r_factor_all_high} 
\end{figure}

\begin{figure}[h]
    \centering
    \begin{subfigure}[b]{0.35\textwidth}
    \caption{\textbf{1.2 keV pnCCD}}
    \includegraphics[scale=0.40]{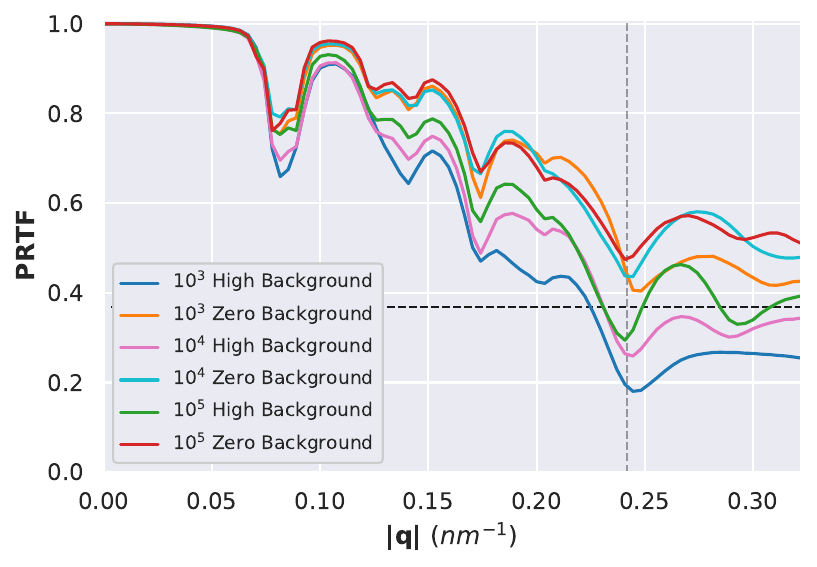}
    \label{prtf_normal_pnccd_1}
    \end{subfigure}
    \centering
    \begin{subfigure}[b]{0.35\textwidth}
    \caption{\textbf{2.5 keV pnCCD}}
    \includegraphics[scale=0.40]{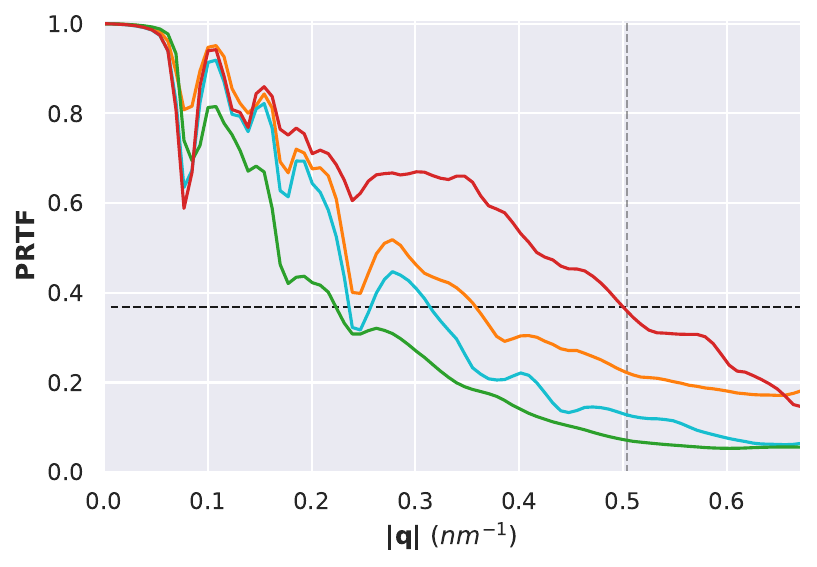}
    \label{prtf_normal_pnccd_2}
    \end{subfigure}
    \centering
    \begin{subfigure}[b]{0.35\textwidth}
    \caption{\textbf{6.0 keV AGIPD}}
    \includegraphics[scale=0.40]{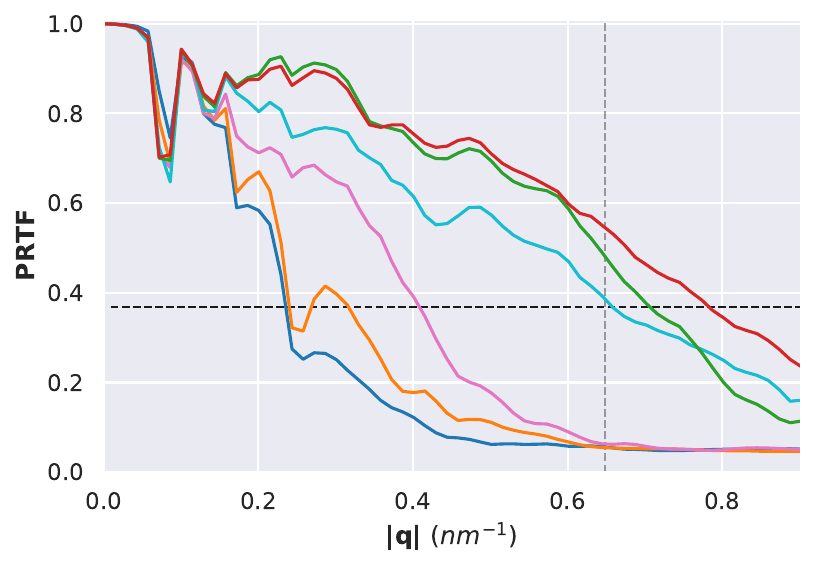}
    \label{prtf_normal_agipd}
    \end{subfigure}
    \caption{PRTF curves for all three energies under high and zero background conditions.}
    \label{prtf_all_high}
\end{figure}

\begin{figure}[h]
    \centering
    \begin{subfigure}[b]{0.35\textwidth}
    \caption{\textbf{1.2 keV pnCCD}}
    \includegraphics[scale=0.40]{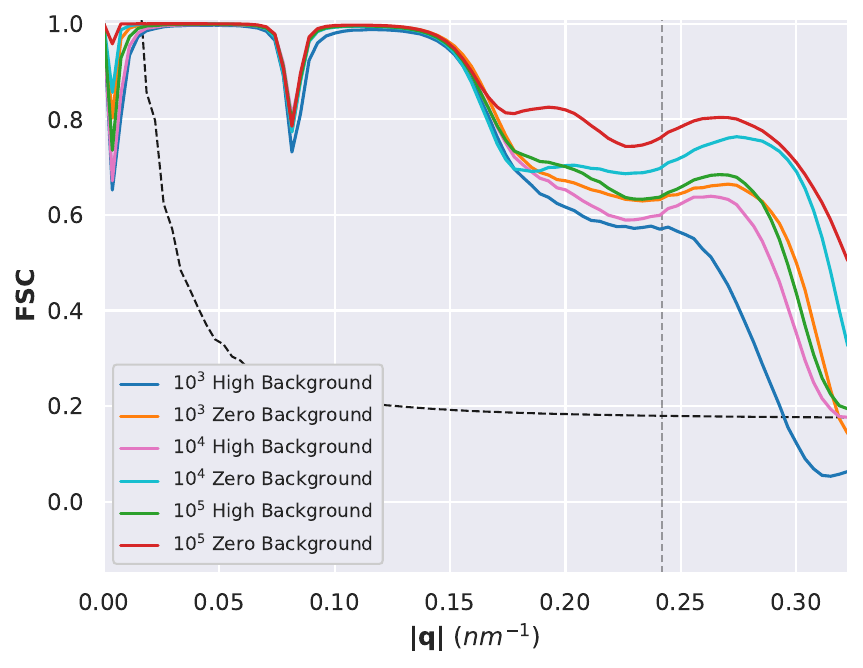}
    \label{fsc_normal_pnccd_1}
    \end{subfigure}
    \centering
    \begin{subfigure}[b]{0.35\textwidth}
    \caption{\textbf{2.5 keV pnCCD}}
    \includegraphics[scale=0.40]{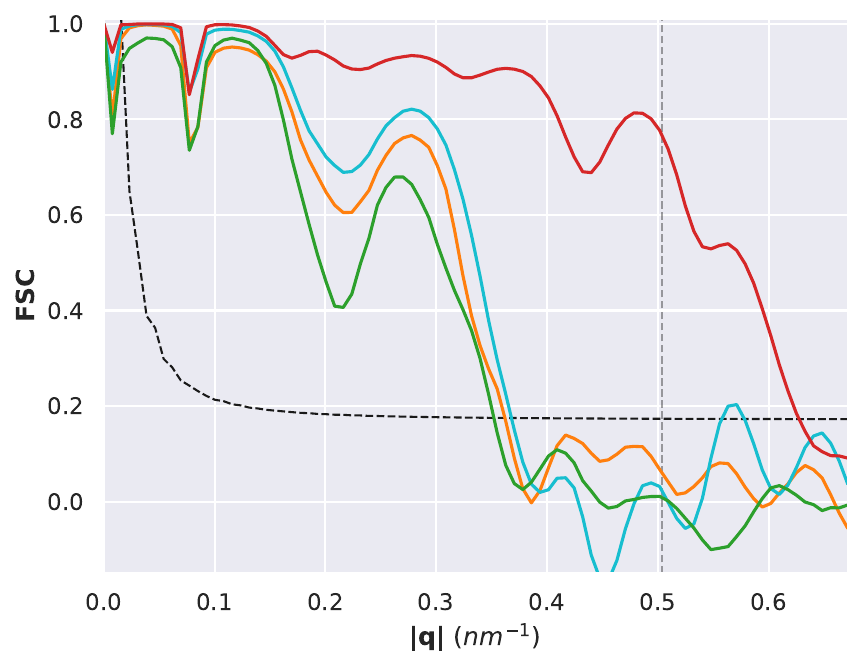}
    \label{fsc_normal_pnccd_2}
    \end{subfigure}
    \centering
    \begin{subfigure}[b]{0.35\textwidth}
    \caption{\textbf{6.0 keV AGIPD}}
    \includegraphics[scale=0.40]{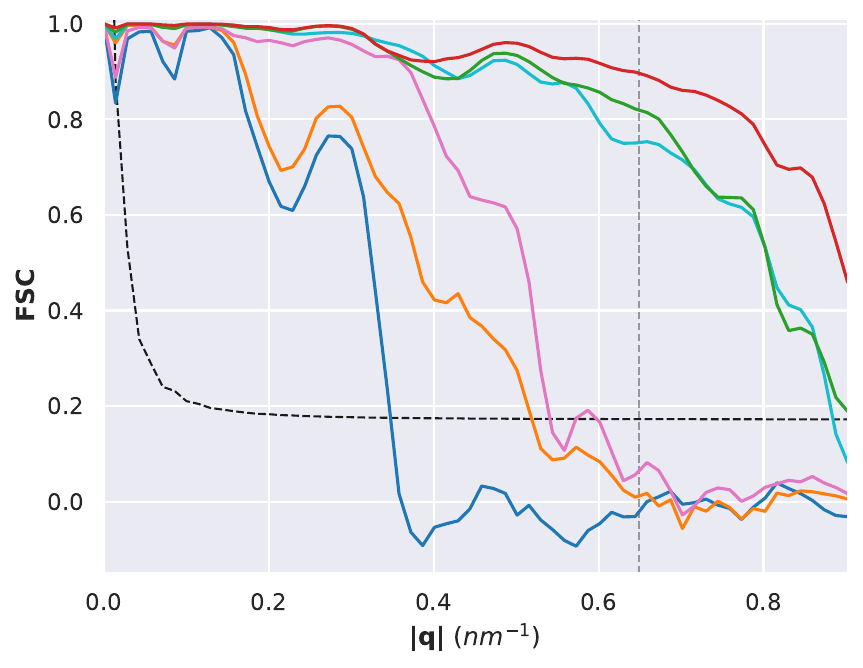}
    \label{fsc_normal_agipd}
    \end{subfigure}
    \caption{Radially averaged FSC curves for high and zero background 
 reconstructions for all three different energies. The half-bit curve is shown as a black dashed line.}
    \label{fsc_all_high}
\end{figure}

\begin{figure}[h]
    \centering
    \begin{subfigure}[b]{0.35\textwidth}
    \caption{\textbf{1.2 keV pnCCD}}
    \includegraphics[scale=0.40]{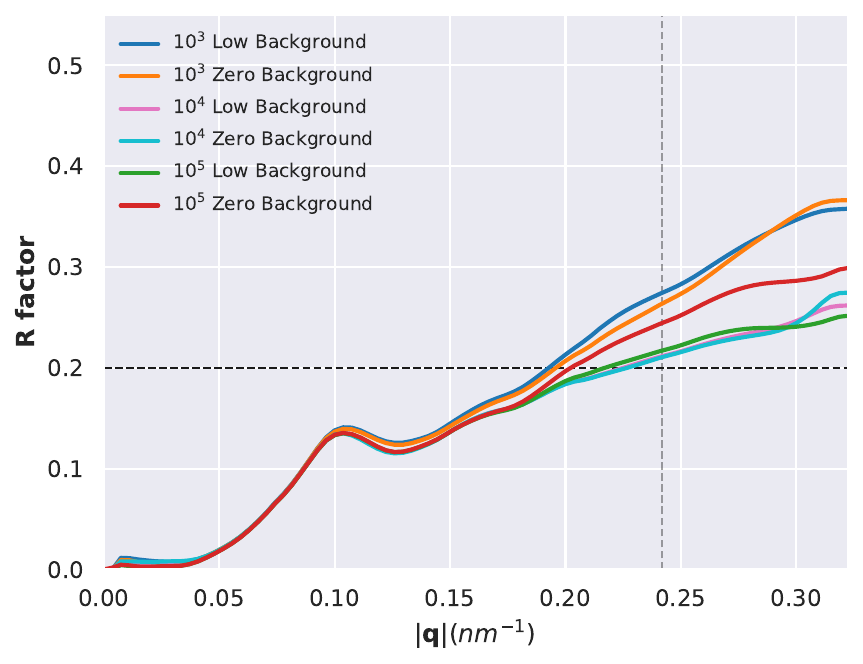}
    \label{rfactor_pnccd_1_low}
    \end{subfigure}
    \centering
    \begin{subfigure}[b]{0.35\textwidth}
    \caption{\textbf{2.5 keV pnCCD}}
    \includegraphics[scale=0.40]{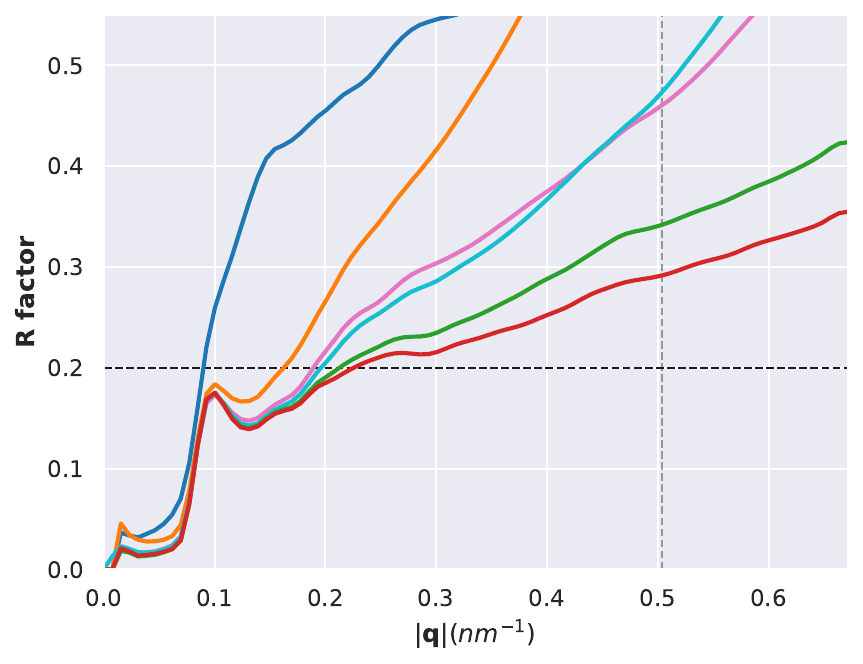}
    \label{rfactor_pnccd_2_low}
    \end{subfigure}
    \centering
    \caption{R factor curves for two energies under low and zero background conditions.}
    \label{r_factor_all_low} 
\end{figure}

\begin{figure}[h]
    \centering
    \begin{subfigure}[b]{0.35\textwidth}
    \caption{\textbf{1.2 keV pnCCD}}
    \includegraphics[scale=0.40]{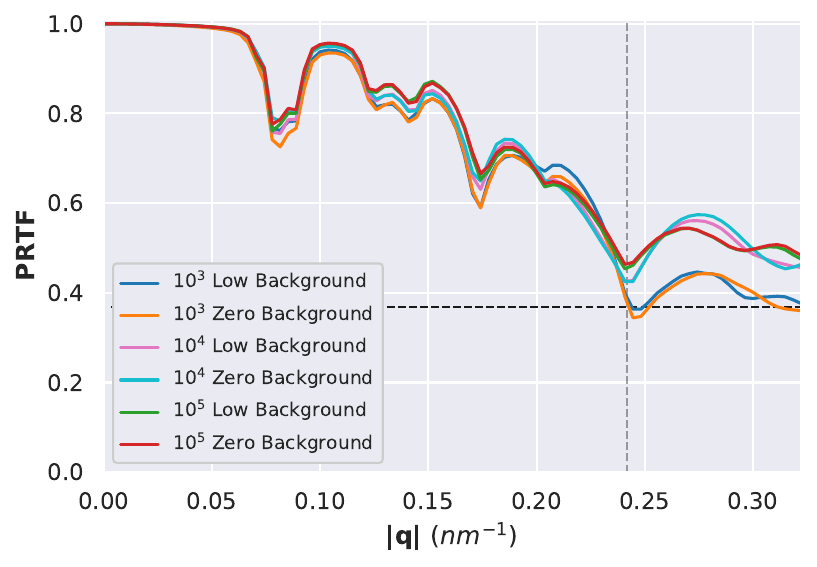}
    \label{prtf_low_pnccd_1}
    \end{subfigure}
    \centering
    \begin{subfigure}[b]{0.35\textwidth}
    \caption{\textbf{2.5 keV pnCCD}}
    \includegraphics[scale=0.40]{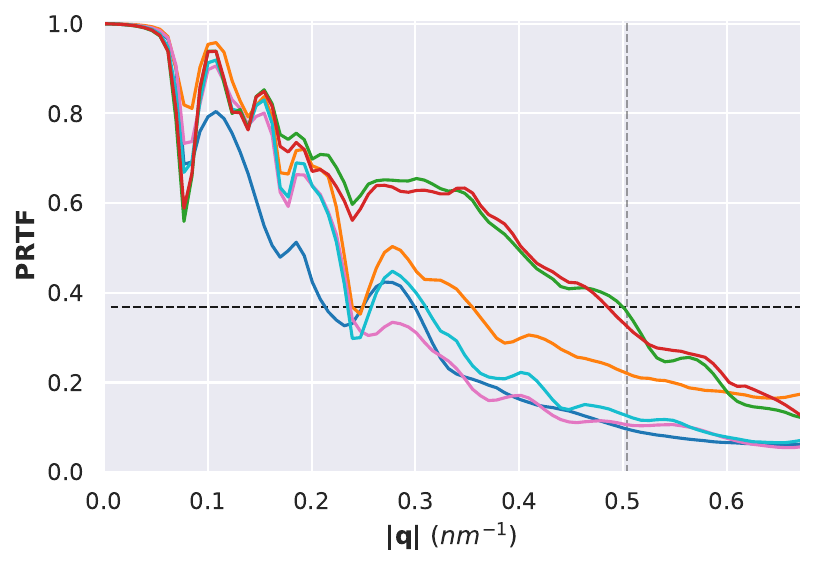}
    \label{prtf_low_pnccd_2}
    \end{subfigure}
    \centering
    \caption{PRTF curves for two energies under low and zero background.}
    \label{prtf_all_low}
\end{figure}

\begin{figure}[h]
    \centering
    \begin{subfigure}[b]{0.35\textwidth}
    \caption{\textbf{1.2 keV pnCCD}}
    \includegraphics[width=0.3\paperwidth]{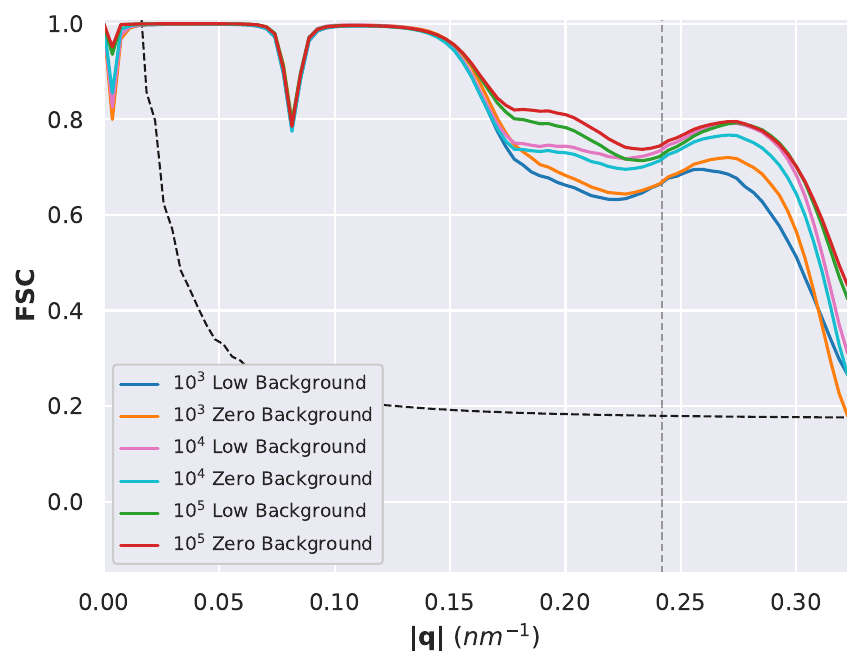}
    \label{fsc_low_pnccd_1}
    \end{subfigure}
    \centering
    \begin{subfigure}[b]{0.35\textwidth}
    \caption{\textbf{2.5 keV pnCCD}}
    \includegraphics[width=0.3\paperwidth]{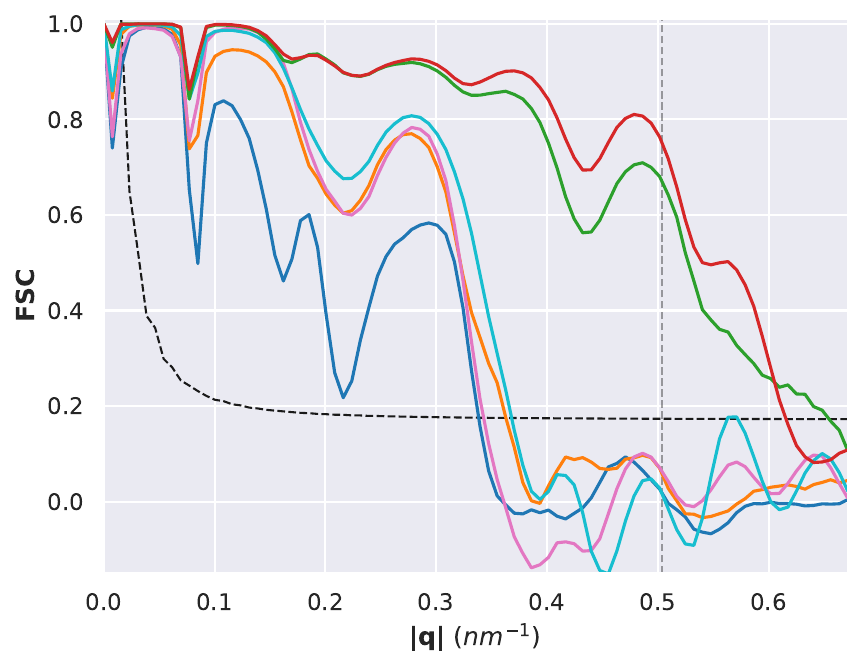}
    \label{fsc_low_pnccd_2}
    \end{subfigure}
    \centering
    \caption{Radially averaged FSC curves for low and zero background for 1.2 keV and 2.5 keV. The half-bit curve is shown as a black dashed line.}
    \label{fsc_all_low}
\end{figure}

\begin{figure}[h]
\centering
\includegraphics[scale=0.43]{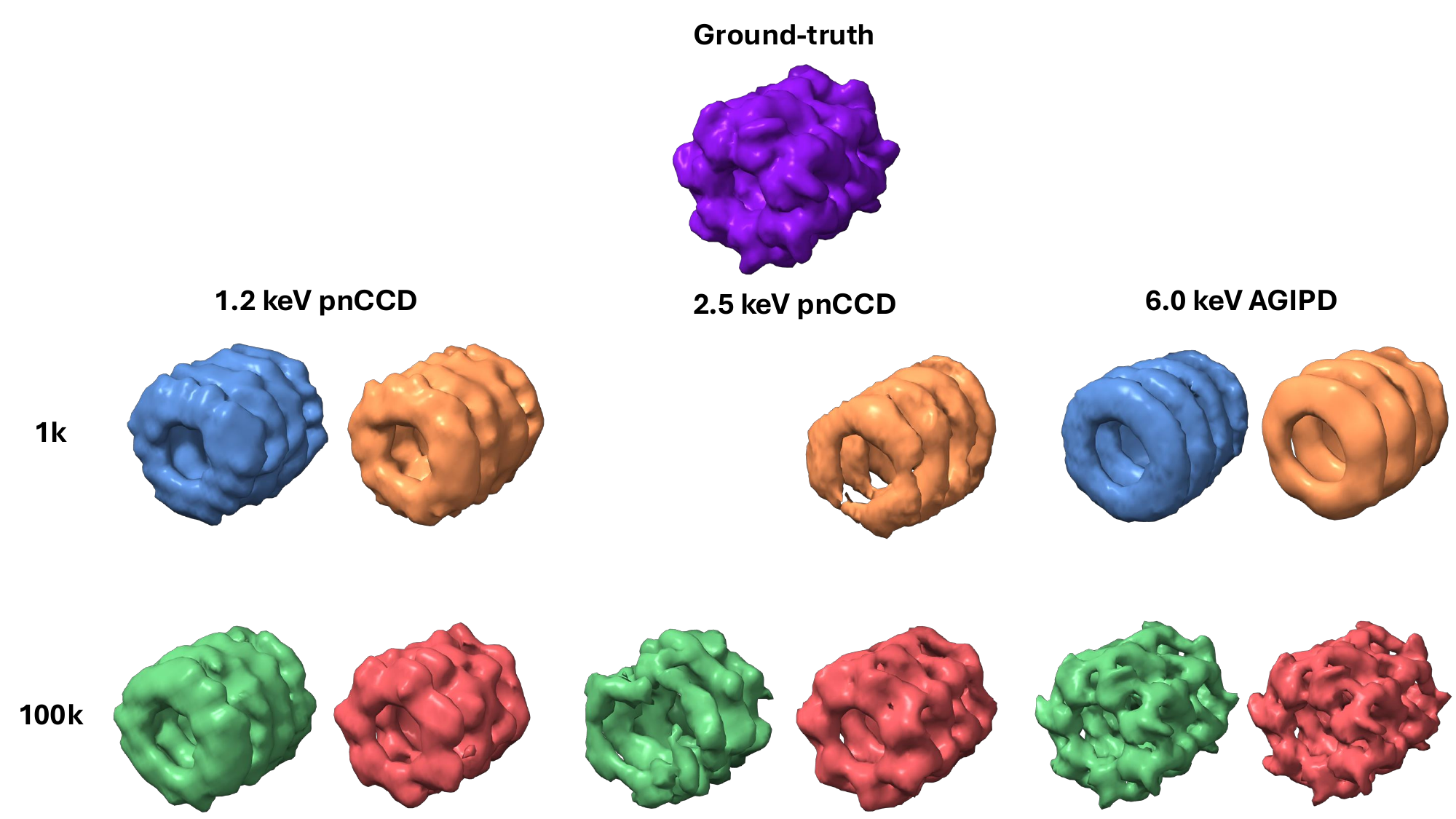}
\caption{Reconstructed electron density for all three geometries for $10^3$, and $10^5$ pattern reconstructions under high (in blue and green on the left) and zero background conditions (in orange and red on the right). The ground truth is shown on top of the reconstructed electron densities. The high background reconstruction at 2.5 keV with $10^3$ patterns failed to converge and is not shown.}
\label{dens_high}
\end{figure}

\begin{figure}[h]
\centering
\includegraphics[scale=0.43]{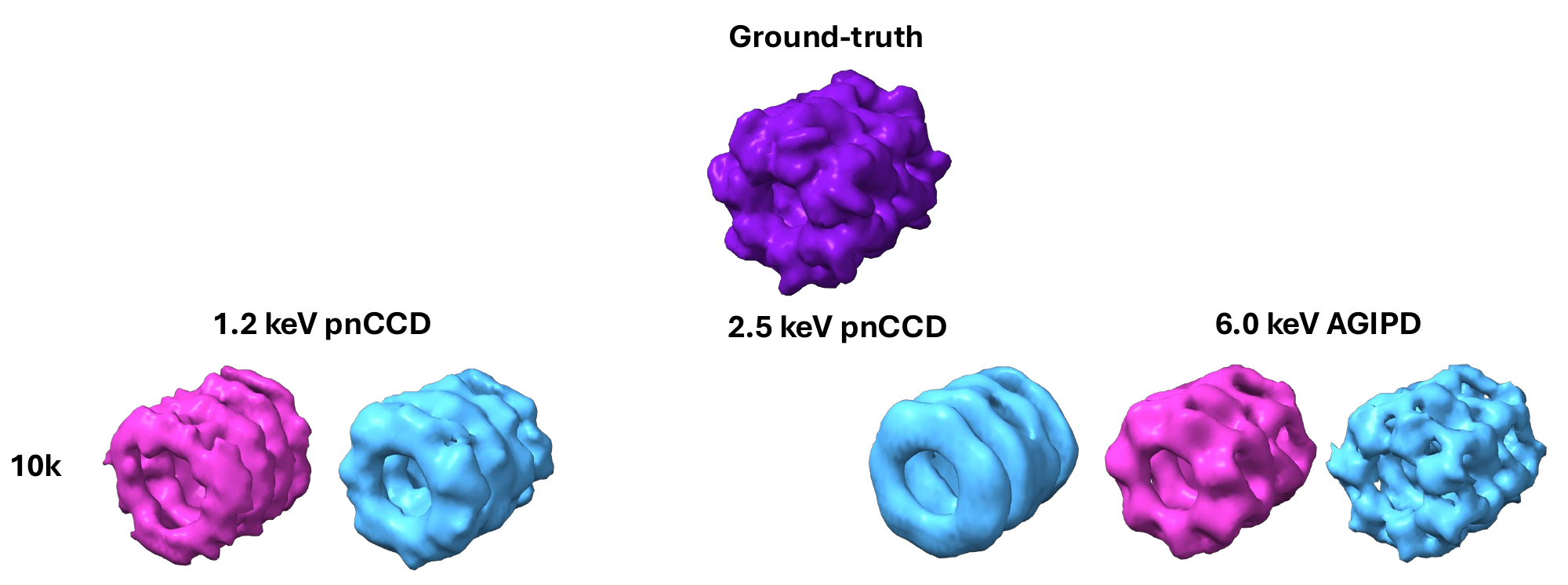}
\caption{Reconstructed electron density for all three geometries for $10^4$ pattern reconstructions under high (in magenta on the left) and zero background conditions (in blue on the right). The ground truth is shown on top of the reconstructed electron densities. The high background reconstruction at 2.5 keV failed to converge and is not shown.}
\label{dens_high_10k}
\end{figure}

\begin{figure}[h]
\centering
\includegraphics[scale=0.43]{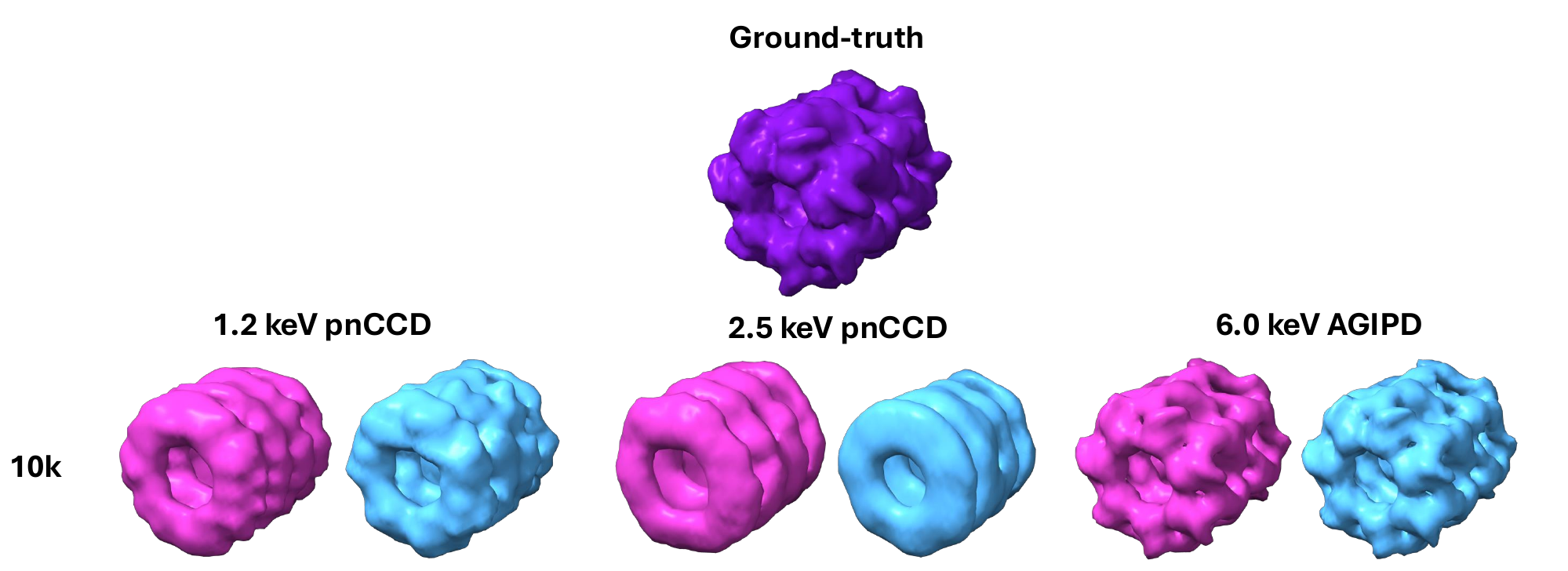}
\caption{Reconstructed electron density for all three energies for $10^4$ pattern reconstructions under medium (in magenta on the left) and zero background conditions (in blue on the right). The ground truth is shown on top of the reconstructed electron densities.}
\label{dens_medium_10k}
\end{figure}

\begin{figure}[h]
\centering
\includegraphics[scale=0.43]{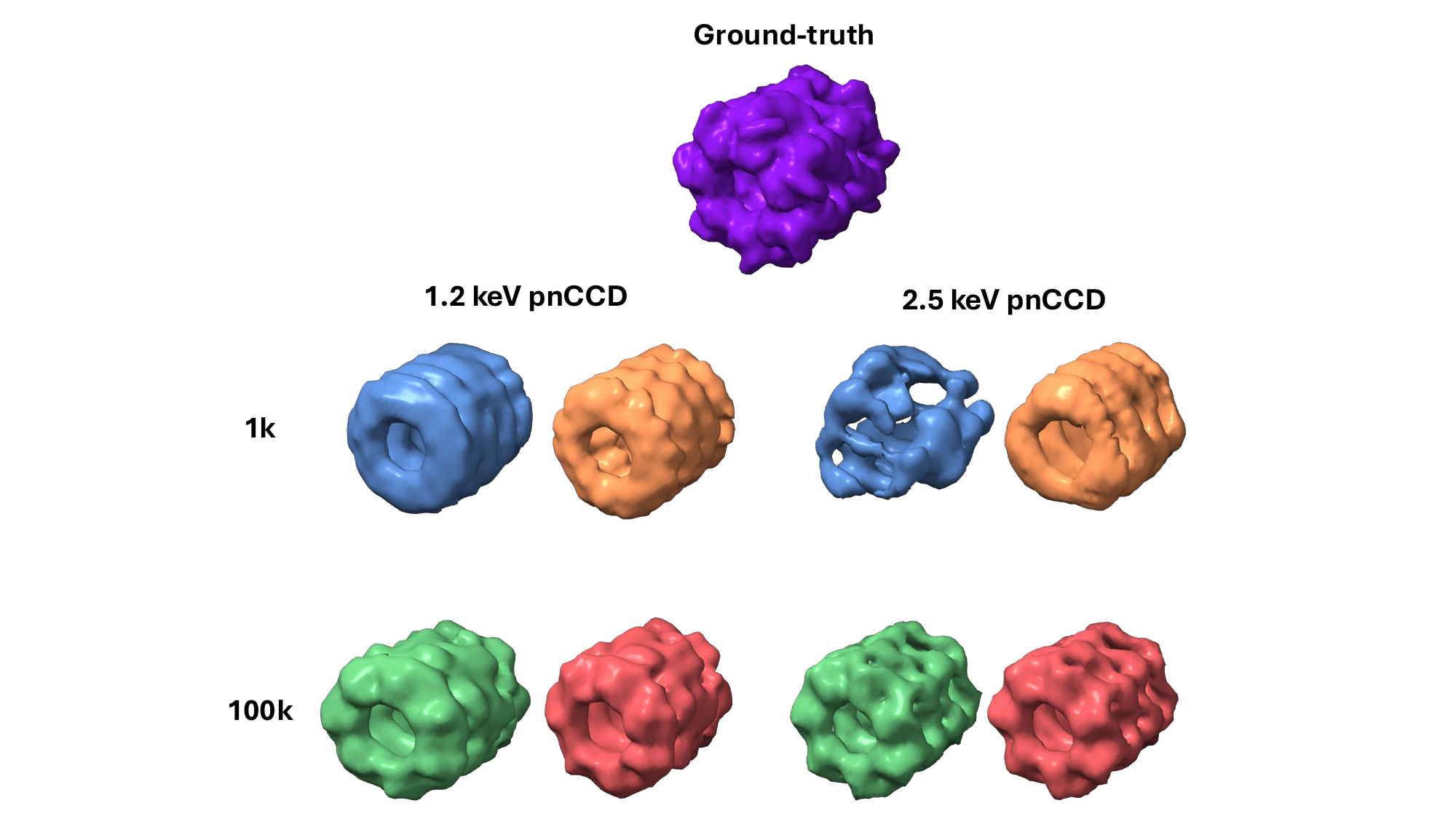}
\caption{Reconstructed electron density for two energies for $10^3$, and $10^5$ pattern reconstructions under low (in blue and green on the left) and zero background conditions (in orange and red on the right). The ground truth is shown on top of the reconstructed electron densities.}
\label{dens_low}
\end{figure}

\begin{figure}[h]
\centering
\includegraphics[scale=0.43]{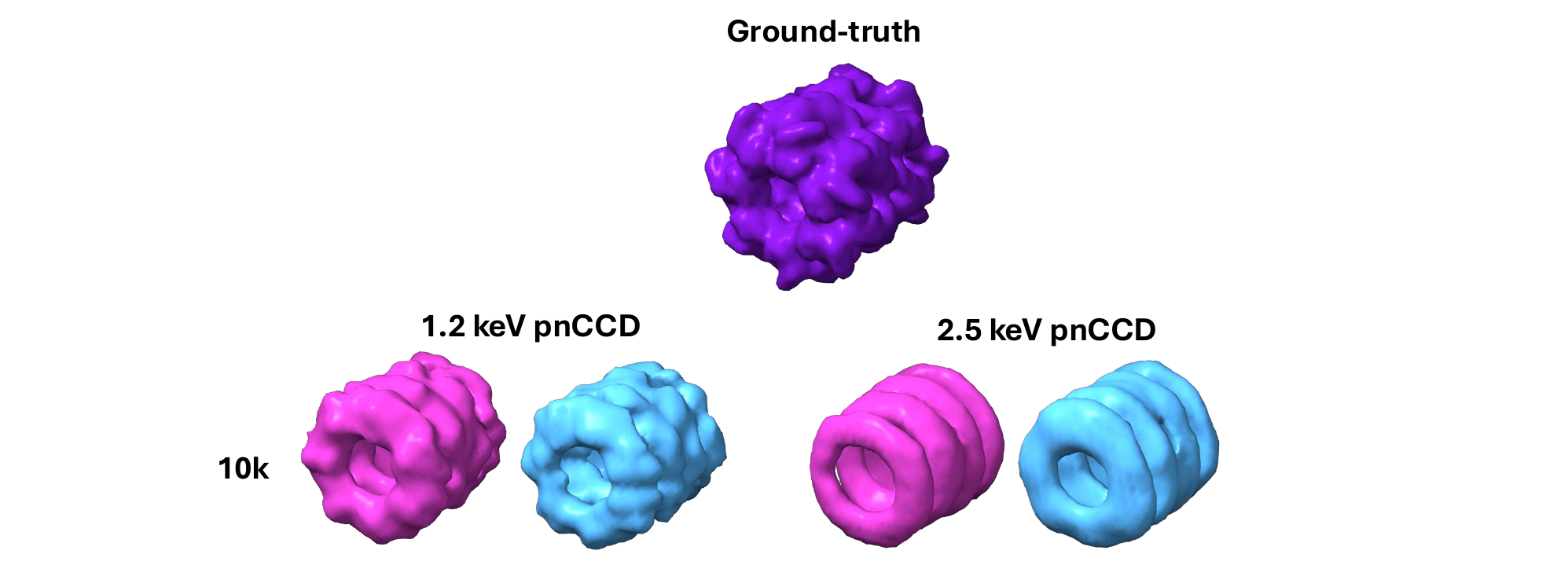}
\caption{Reconstructed electron density for two energies for $10^4$ pattern reconstructions under low (in magenta on the left) and zero (in blue on the right) background conditions. The ground truth is shown on top of the reconstructed electron densities.}
\label{dens_low_10k}
\end{figure}

\newpage
\begin{figure}[h]
    \begin{subfigure}[b]{1.0\textwidth}
    \centering
    \caption{\textbf{Condor}}
    \includegraphics[scale=0.45]{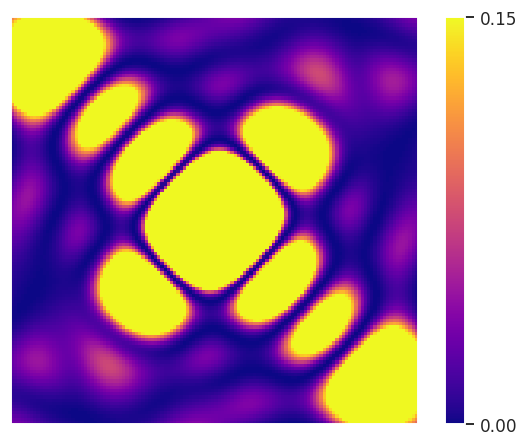}
    \label{emc_gt_pnccd_1}
    \end{subfigure}
    \centering
    \begin{subfigure}[b]{0.3\textwidth}
    \caption{\textbf{High}}
    \includegraphics[scale=0.45]{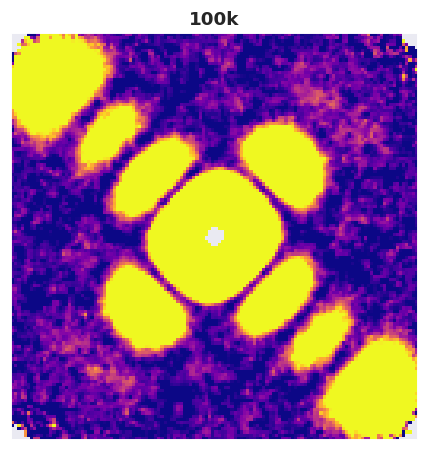}
    \label{emc_pnccd_1_1}
    \end{subfigure}
    \centering
    \begin{subfigure}[b]{0.3\textwidth}
    \caption{\textbf{Medium}}
    \includegraphics[scale=0.45]{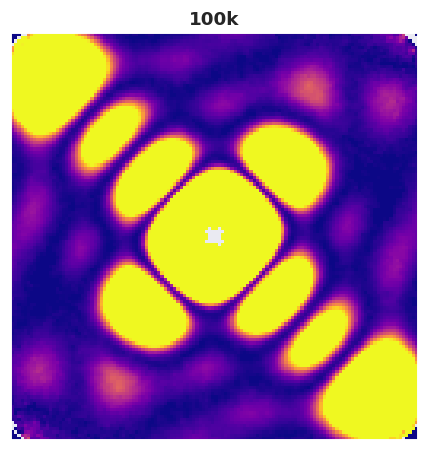}
    \label{emc_pnccd_1_2}
    \end{subfigure}
    \centering
    \begin{subfigure}[b]{0.3\textwidth}
    \caption{\textbf{Low}}
    \includegraphics[scale=0.45]{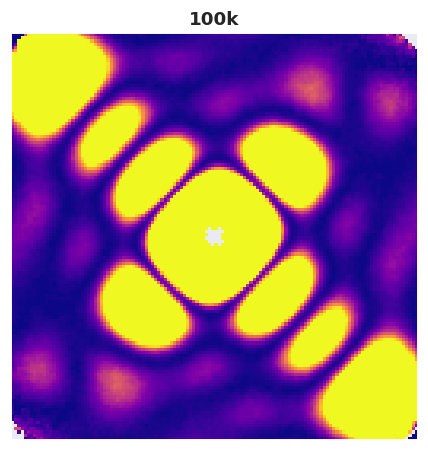}
    \label{emc_pnccd_1_3}
    \end{subfigure}
    \centering
    \begin{subfigure}[b]{0.3\textwidth}
    \includegraphics[scale=0.45]{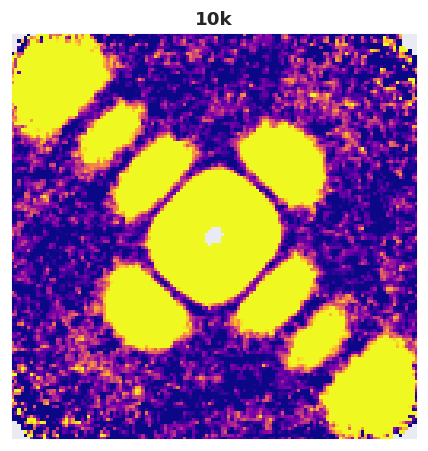}
    \label{emc_pnccd_1_4}
    \end{subfigure}
    \centering
    \begin{subfigure}[b]{0.3\textwidth}
    \includegraphics[scale=0.45]{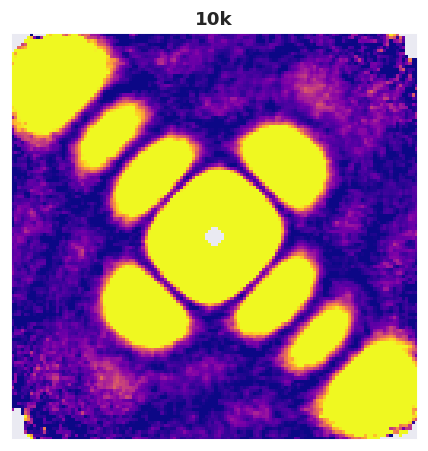}
    \label{emc_pnccd_1_5}
    \end{subfigure}
    \centering
    \begin{subfigure}[b]{0.3\textwidth}
    \includegraphics[scale=0.45]{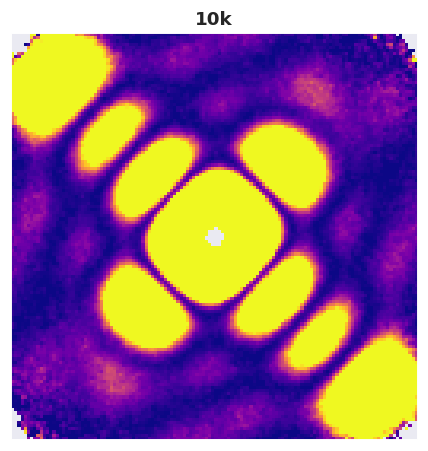}
    \label{emc_pnccd_1_6}
    \end{subfigure}
    \centering
    \begin{subfigure}[b]{0.3\textwidth}
    \includegraphics[scale=0.45]{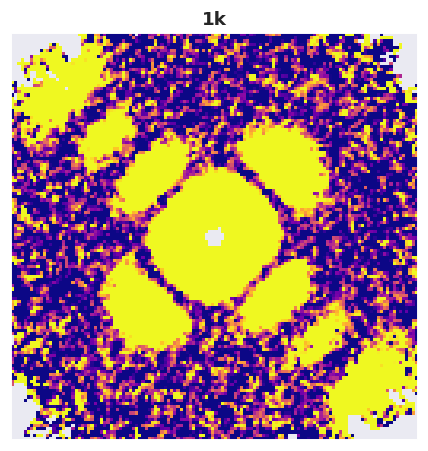}
    \label{emc_pnccd_1_7}
    \end{subfigure}
    \centering
    \begin{subfigure}[b]{0.3\textwidth}
    \includegraphics[scale=0.45]{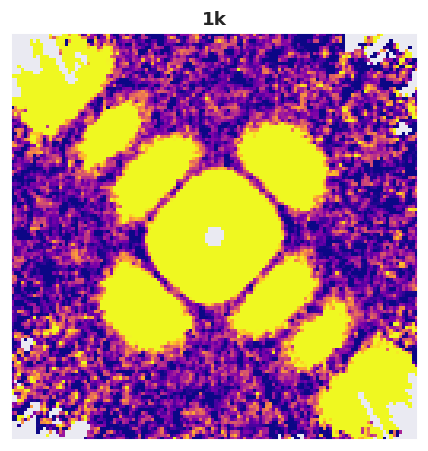}
    \label{emc_pnccd_1_8}
    \end{subfigure}
    \centering
    \begin{subfigure}[b]{0.3\textwidth}
    \includegraphics[scale=0.45]{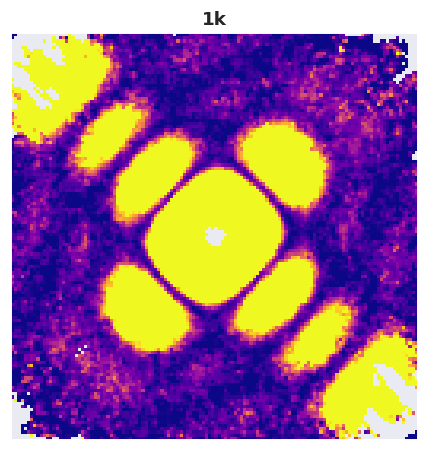}
    \label{emc_pnccd_1_9}
    \end{subfigure}
    \caption{Central XY slices of the Condor ground truth and aligned EMC reconstructions for each background level for 1.2 keV reconstructions. The color map representing intensity has arbitrary units. Each plot has the same minimum and maximum intensity range.}
    \label{emc_slices_pnccd_1} 
\end{figure}
\newpage
\begin{figure}[h]
    \begin{subfigure}[b]{1.0\textwidth}
    \centering
    \caption{\textbf{Condor}}
    \includegraphics[scale=0.45]{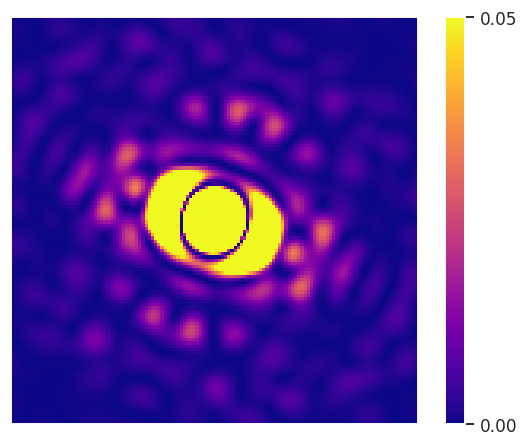}
    \label{emc_gt_pnccd_2}
    \end{subfigure}
    \centering
    \begin{subfigure}[b]{0.3\textwidth}
    \caption{\textbf{High}}
    \includegraphics[scale=0.45]{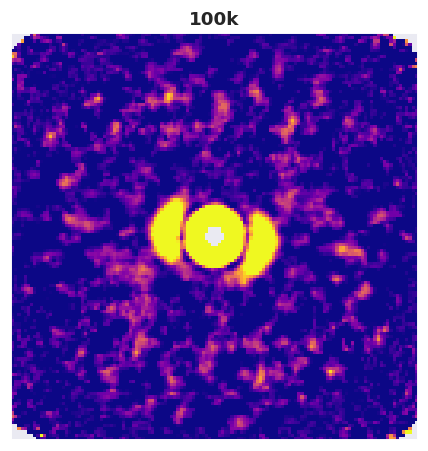}
    \label{emc_pnccd_2_1}
    \end{subfigure}
    \centering
    \begin{subfigure}[b]{0.3\textwidth}
    \caption{\textbf{Medium}}
    \includegraphics[scale=0.45]{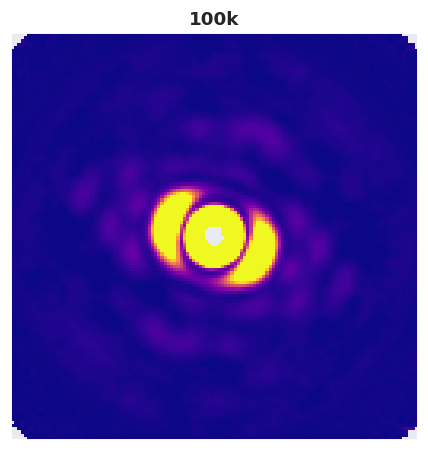}
    \label{emc_pnccd_2_2}
    \end{subfigure}
    \centering
    \begin{subfigure}[b]{0.3\textwidth}
    \caption{\textbf{Low}}
    \includegraphics[scale=0.45]{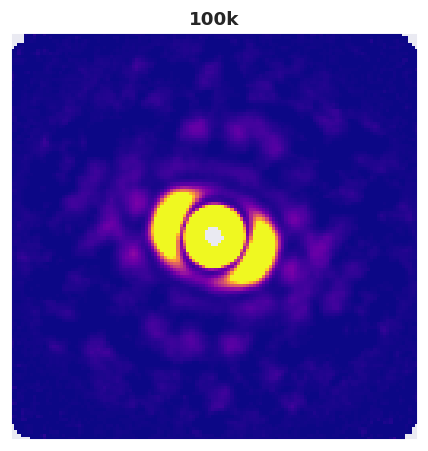}
    \label{emc_pnccd_2_3}
    \end{subfigure}
    \centering
    \begin{subfigure}[b]{0.3\textwidth}
    \phantom{\includegraphics[scale=0.35]{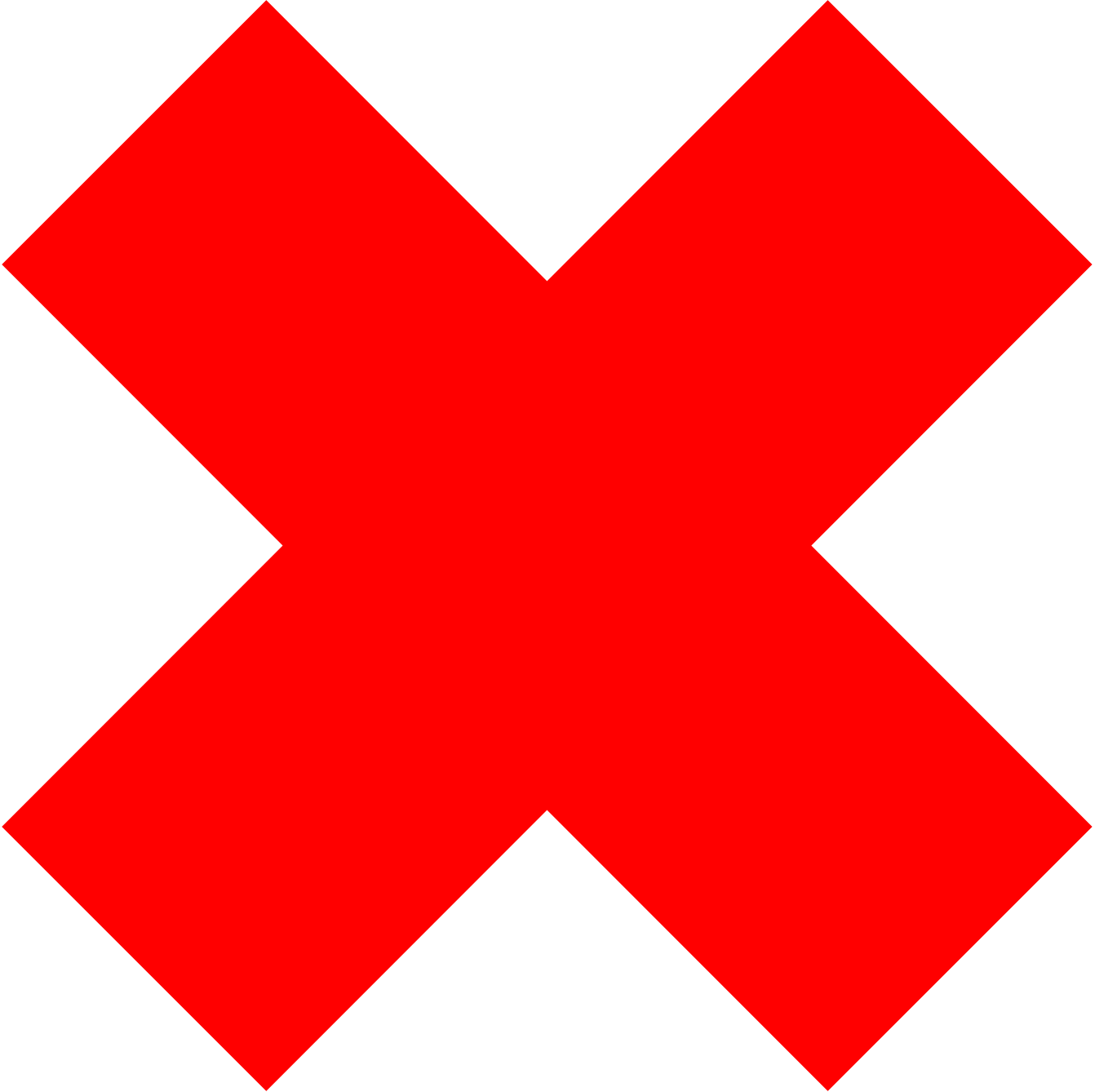}}
    \label{emc_pnccd_2_4}
    \end{subfigure}
    \centering
    \begin{subfigure}[b]{0.3\textwidth}
    \includegraphics[scale=0.45]{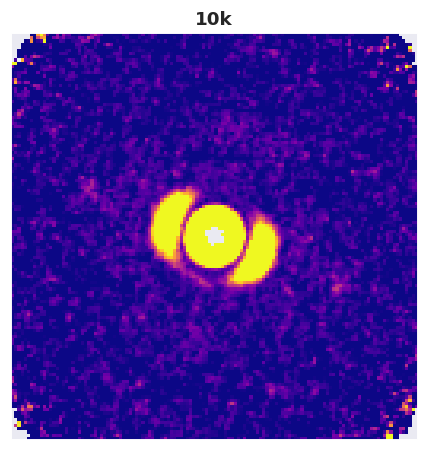}
    \label{emc_pnccd_2_5}
    \end{subfigure}
    \centering
    \begin{subfigure}[b]{0.3\textwidth}
    \includegraphics[scale=0.45]{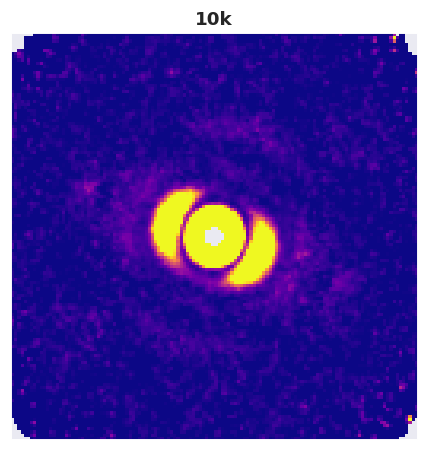}
    \label{emc_pnccd_2_6}
    \end{subfigure}
    \centering
    \begin{subfigure}[b]{0.3\textwidth}
    \phantom{\includegraphics[scale=0.45]{figures/cross.png}}
    \label{emc_pnccd_2_7}
    \end{subfigure}
    \centering
    \begin{subfigure}[b]{0.3\textwidth}
    \includegraphics[scale=0.45]{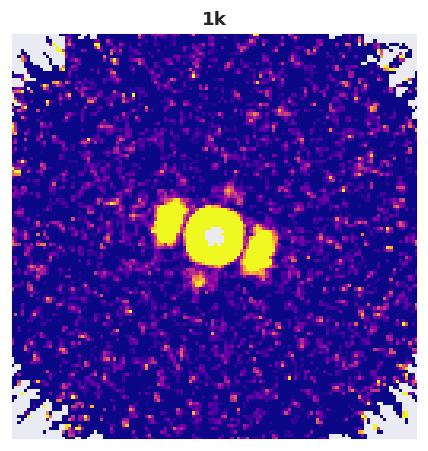}
    \label{emc_pnccd_2_8}
    \end{subfigure}
    \centering
    \begin{subfigure}[b]{0.3\textwidth}
    \includegraphics[scale=0.45]{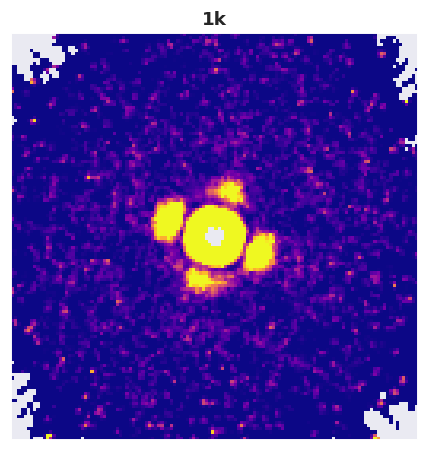}
    \label{emc_pnccd_2_9}
    \end{subfigure}
    \caption{Central XY slices of the Condor ground truth and aligned EMC reconstructions for each background level for 2.5 keV reconstructions. Missing reconstructions are represented by a blank spot in the figure. The color map representing intensity has arbitrary units. Each plot has the same minimum and maximum intensity range.}
    \label{emc_slices_pnccd_2} 
\end{figure}
\newpage
\begin{figure}[h]
    \begin{subfigure}[b]{1.0\textwidth}
    \centering
    \caption{\textbf{Condor}}
    \includegraphics[scale=0.5]{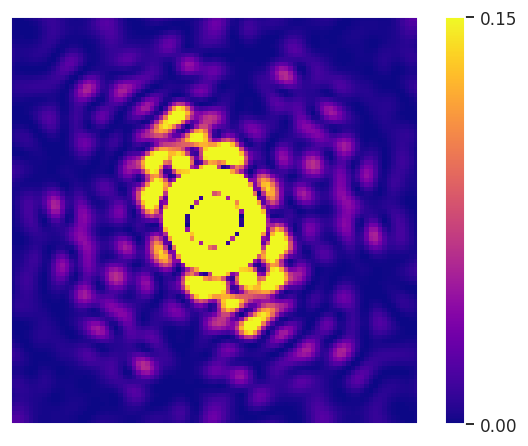}
    \label{emc_gt_agipd}
    \end{subfigure}
    \centering
    \begin{subfigure}[b]{0.35\textwidth}
    \caption{\textbf{High}}
    \includegraphics[scale=0.5]{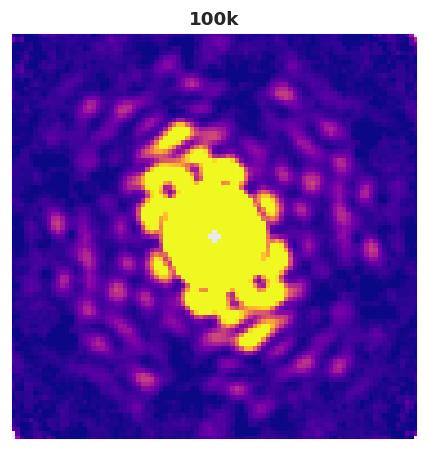}
    \label{emc_agipd_1}
    \end{subfigure}
    \centering
    \begin{subfigure}[b]{0.35\textwidth}
    \caption{\textbf{Medium}}
    \includegraphics[scale=0.5]{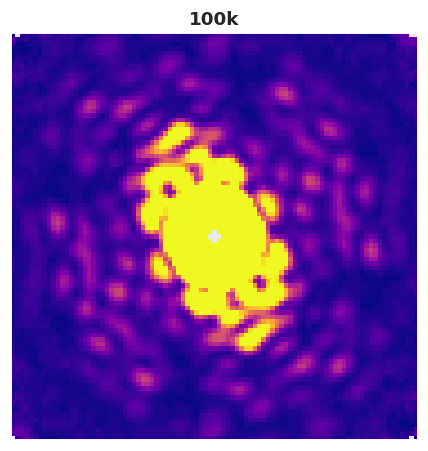}
    \label{emc_agipd_2}
    \end{subfigure}
    \centering
    \begin{subfigure}[b]{0.35\textwidth}
    \includegraphics[scale=0.5]{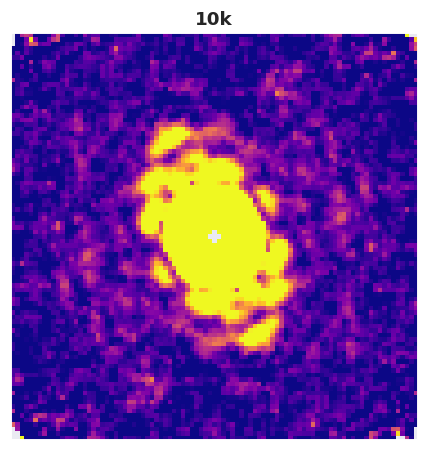}
    \label{emc_agipd_3}
    \end{subfigure}
    \centering
    \begin{subfigure}[b]{0.35\textwidth}
    \includegraphics[scale=0.5]{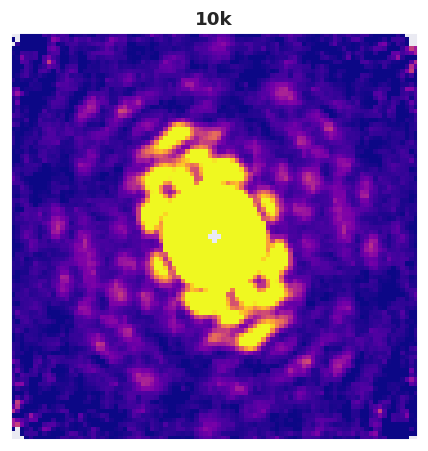}
    \label{emc_agipd_4}
    \end{subfigure}
    \centering
    \begin{subfigure}[b]{0.35\textwidth}
    \includegraphics[scale=0.5]{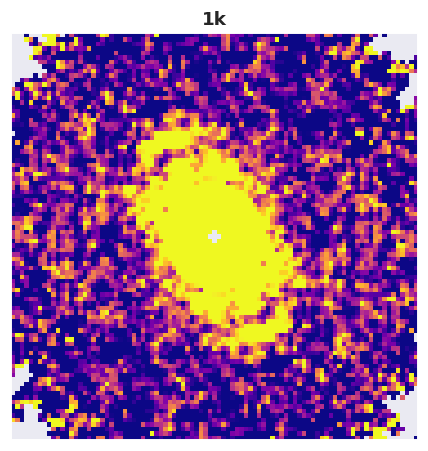}
    \label{emc_agipd_5}
    \end{subfigure}
    \centering
    \begin{subfigure}[b]{0.35\textwidth}
    \includegraphics[scale=0.5]{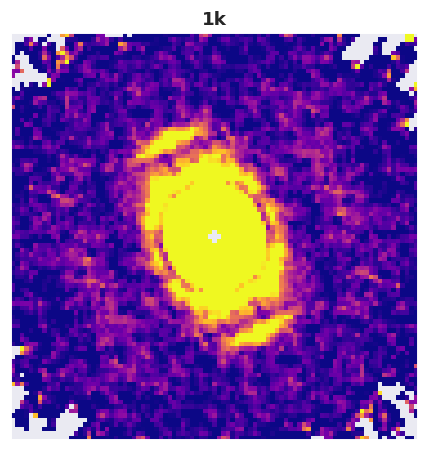}
    \label{emc_agipd_6}
    \end{subfigure}
    \caption{Central XY slices of the Condor ground truth and aligned EMC reconstructions for each background level for 6.0 keV reconstructions. The color map representing intensity has arbitrary units. Each plot has the same minimum and maximum intensity range.}
    \label{emc_slices_agipd}
\end{figure}

\end{document}